\documentclass{aa}  

\usepackage{graphicx}
\usepackage{threeparttable} %for table note
\usepackage{array}
\usepackage[varg]{txfonts}
\usepackage{chemmacros}
\usepackage{caption}
\usepackage{amsmath}

\begin{document}

	%\title{Detection of Fe, Ti and a temperature inversion on the \\dayside of the ultra-hot Jupiter MASCARA-1b}
	
	\title{Detection of Fe and Ti on the dayside of the ultrahot Jupiter MASCARA-1b with CARMENES}
	
	% \title{Hahaha}
	
	\author{B. Guo \inst{1,2}
		\and
		F. Yan \inst{1,2}
		\and
		L.~Nortmann\inst{2}
		\and
		D. Cont\inst{3,4}
		\and
		A. Reiners\inst{2}
		\and
		E. Pall\'e\inst{5,6}
		\and
		D. Shulyak\inst{7}
		\and
		K. Molaverdikhani\inst{3,8,9}
		\and
		Th.~Henning\inst{8}
		\and
		G.~Chen\inst{10,11}
		\and
		M.~Stangret\inst{12}
		\and
		S.~Czesla\inst{13,14}
		\and
		F.~Lesjak\inst{2}
		\and
		M. L\'opez-Puertas\inst{7}         
		\and
		I.~Ribas\inst{15,16},  A.~Quirrenbach\inst{9}, J.~A.~Caballero\inst{17}, P.~J.~Amado\inst{7}
		\and
		M.~Blazek\inst{18}, D.~Montes\inst{19}, J.~C.~Morales\inst{15,16}
		\and E.~Nagel\inst{2}
		\and M.~R.~Zapatero~Osorio\inst{17} 
		%\and M.~Zechmeister\inst{1}, D.~Galad\'i-Enr\'iquez\inst{17},G.~Morello\inst{2}, L.~M.~Lara\inst{5}, D.~Montes\inst{18}
	}
	
	\institute{Department of Astronomy, University of Science and Technology of China, Hefei 230026, China\\
		\email{yanfei@ustc.edu.cn}
		\and
		Institut f\"ur Astrophysik und Geophysik, Georg-August-Universit\"at, Friedrich-Hund-Platz 1, 37077 G\"ottingen, Germany
		\and
		Ludwig-Maximilians-Universit\"at, Universitäts-Sternwarte München, Scheinerstr. 1, 81679, Munich, Germany
		\and
		Exzellenzcluster Origins, Boltzmannstra{\ss}e 2, 85748 Garching, Germany
		\and
		Instituto de Astrof{\'i}sica de Canarias (IAC), V{\'i}a Lactea s/n, 38200 La Laguna, Tenerife, Spain
		\and
		Departamento de Astrof{\'i}sica, Universidad de La Laguna, 38026  La Laguna, Tenerife, Spain
		\and
		Instituto de Astrof\'{\i}sica de Andaluc\'{\i}a - CSIC, Glorieta de la Astronom\'{\i}a s/n, 18008 Granada, Spain
		\and
		Max-Planck-Institut f{\"u}r Astronomie, K{\"o}nigstuhl 17, 69117 Heidelberg, Germany
		\and
		Landessternwarte, Zentrum f\"ur Astronomie der Universit\"at Heidelberg, K\"onigstuhl 12, 69117 Heidelberg, Germany
		\and
		CAS Key Laboratory of Planetary Sciences, Purple Mountain Observatory, Chinese Academy of Sciences, Nanjing 210023, China
		\and    
		CAS Center for Excellence in Comparative Planetology, Hefei 230026, China
		\and
		INAF – Osservatorio Astronomico di Padova, Vicolo dell'Osservatorio 5, 35122, Padova, Italy
		\and
		Th{\"u}ringer Landessternwarte Tautenburg, Sternwarte 5, 07778 Tautenburg, Germany
		\and
		Hamburger Sternwarte, Universit{\"a}t Hamburg, Gojenbergsweg 112, 21029 Hamburg, Germany
		\and
		Institut de Ci\`encies de l'Espai (CSIC-IEEC), Campus UAB, c/ de Can Magrans s/n, 08193 Bellaterra, Barcelona, Spain
		\and
		Institut d'Estudis Espacials de Catalunya (IEEC), 08034 Barcelona, Spain
		\and
		Centro de Astrobiolog\'{i}a (CSIC-INTA), ESAC, Camino Bajo del Castillo s/n, E-28692 Villanueva de la Ca\~nada, Madrid, Spain
		\and
		Centro Astron{\'o}nomico Hispano-Alem{\'a}n (CSIC--Junta de Andaluc\'ia), Observatorio Astron{\'o}nomico de Calar Alto, Sierra de los Filabres, 04550 G{\'e}rgal, Almer\'ia, Spain
		\and
		Departamento de F\'{i}sica de la Tierra y Astrof\'{i}sica and IPARCOS-UCM (Instituto de F\'{i}sica de Part\'{i}culas y del Cosmos de la UCM), Facultad de Ciencias F\'{i}sicas, Universidad Complutense de Madrid, 28040, Madrid, Spain
		% \and
		%Centro de Astrobiolog\'{i}a (CSIC-INTA), Carretera de Ajalvir, km 4, 28850 Torrej\'{o}n de Ardoz, Madrid, Spain
	}
	
	\date{Received --; accepted --}

	\abstract
	{Ultrahot Jupiters are a type of gaseous exoplanet that orbit extremely close to their host star, resulting in significantly high equilibrium temperatures. In recent years, high-resolution emission spectroscopy has been broadly employed in observing the atmospheres of ultrahot Jupiters. We used the CARMENES spectrograph to observe the high-resolution spectra of the dayside hemisphere of MASCARA-1b in both visible and near-infrared. Through cross-correlation analysis, we detected signals of \ion{Fe}{i} and \ion{Ti}{i}. Based on these detections, we conducted an atmospheric retrieval and discovered the presence of a strong inversion layer in the planet's atmosphere. 
		The retrieved Ti and Fe abundances are broadly consistent with solar abundances. In particular, we obtained a relative abundance of [Ti/Fe] as $-1.0 \pm 0.8$ under the free retrieval and $-0.4^{+0.5}_{-0.8}$ under the chemical equilibrium retrieval, suggesting the absence of significant titanium depletion on this planet. Furthermore, we considered the influence of planetary rotation on spectral line profiles. The resulting equatorial rotation speed was determined to be $4.4^{+1.6}_{-2.0}\,\mathrm{km\,s^{-1}}$, which agrees with the rotation speed induced by tidal locking. }
	% We also proposed two possible reasons for the discrepancies between the retrieved results with fixed volume mixing ratios of \ion{Fe}{i} or atmospheric chemical equilibrium: non-local thermodynamic equilibrium (NLTE) effects and transport-induced quenching. 
	
	
	\titlerunning{Fe and Ti in MASCARA-1b}
	\maketitle
	
	%-------------------------------------------------------------------
	
	\section{Introduction}
	
	Ultrahot Jupiters (UHJs) are giant gas planets that are exceptionally close to their host stars, which results in very high equilibrium temperatures ($T_\mathrm{eq} > 2000\,\mathrm{K}$). These planets generally exhibit a temperature inversion on their dayside. This is due to the strong absorption of radiation from the star by certain chemical species in the atmosphere, such as iron, hydrogen, TiO, and VO \citep[e.g.,][]{inversions_1, inversions_2,inversions_3, inversions_4, NLTE}, which leads to an increase in temperature in the upper atmospheric layers.
	
	Over the past several years, numerous studies have been conducted to investigate the atmospheres of UHJs using high-resolution emission spectroscopy. For example, in the dayside atmosphere of KELT-20b/MASCARA-2b, chemical species like \ion{Fe}{i}, \ion{Si}{i}, \ion{Fe}{ii}, \ion{Cr}{i}, and \ion{Ni}{i} have been detected \citep[e.g.,][]{KELT20b_detection, KELT-20b_introduction1, KELT-20b_introduction2, KELT-20b_introduction3, KELT-20b_introduction4}. Similarly, in the case of KELT-9b, \ion{Fe}{i}, \ion{Mg}{i}, \ion{Si}{i}, and \ion{Ca}{ii} have been successfully detected using high-resolution spectrometers such as HARPS-N, CARMENES, and MAROON-X \citep[e.g.,][]{KELT-9b_introduction2, KELT-9b_introduction3, KELT-9b_introduction4, KELT-9b_introduction1}.
	\ion{Fe}{i}, \ion{Si}{i}, \ion{V}{i}, \ion{Ti}{i}, CO, OH, and TiO have been detected in WASP-33b \citep[e.g.,][]{WASP-33b_introduction2, WASP-33b_introduction3, WASP-33b_introduction4, WASP-33b_introduction5, WASP-33b_introduction6, KELT-20b_introduction1, WASP-33b_introduction1, WASP-33b_introduction7}, and the latest research has measured the mixing ratios of CO, \ch{H2O}, and OH, indicating that \ch{H2O} on the dayside is almost completely photodissociated \citep{WASP-33b_mass-mixing_ratio}. 
	In addition, the dayside atmospheres of WASP-18b, WASP-76b, WASP-189b, and WASP-121b have also been probed with high-resolution spectroscopy \citep[e.g.][]{new_retrieval_method, WASP-121b-cold_trap, WASP-18b_retrieval}.
	
	In this paper, we present the detection of \ion{Fe}{i} and \ion{Ti}{i} in the thermal emission spectrum from the dayside of \object{MASCARA-1b} along with the retrieval of the atmospheric properties. The planet \object{MASCARA-1b} was discovered by \citet{MASCARA_parameters} and orbits a bright and rapidly rotating A8-type star with a period of 2.18 days. With a mass of $3.7\,M_{\mathrm{J}}$ and a radius of $1.5\,R_{\mathrm{J}}$, the planet has a density similar to that of Jupiter. It has one of the highest densities of the known UHJs. Furthermore, this planet exhibits a surface equilibrium temperature of $T_\mathrm{eq} = 2594.3^{+1.6}_{-1.5}\,\mathrm{K}$ \citep{spi}, placing it among the hottest and most radiative exoplanets known.
	
	There are several previous studies on the atmosphere of \object{MASCARA-1b}. Spitzer and CHEOPS observations reveal that \object{MASCARA-1b} has a near-polar orbit and uncovers a hint of dayside reflection \citep{spi}. Some previous studies have also used ESPRESSO \citep{espresso} and HARPS-N \citep{harps} to observe its transmission spectrum, but unfortunately no atmospheric signals were detected. This may be due to its high surface gravity ($\mathrm{log}{\,}\textit{g} = 3.63\,\mathrm{cgs}$), which results in low scale height, thereby limiting the applicability of high-resolution transmission spectroscopy to \object{MASCARA-1b}. However, due to the high integrated dayside temperature of the planet ($T_{\mathrm{day}} = 3062^{+66}_{-68}\,\mathrm{K}$; \citealt{spi}) and the brightness of the host star, we can better study its atmosphere through emission spectroscopy. The PEPSI Exoplanet Transit Survey detected signals of \ion{Fe}{i}, \ion{Ti}{i}, and \ion{Cr}{i} \citep{MASCARA-1b_CrI} in its dayside atmosphere, while the high-resolution infrared spectroscopy of CRIRES+ detected signals of CO, \ch{H2O}, and \ion{Fe}{i}, confirming the consistency of C/O with the solar value \citep{CRIRES+_MASACRA1b}.
	
	The paper is organized as follows. In Sect.\ref{obs}, we describe the observations and data reduction. In Sect.\ref{method}, we present the method of atmospheric detection. In Sect.\ref{result}, we show the detection results. Our atmosphere retrieval method and results are described in Sect.\ref{retrieval}. The conclusion is presented in Sect.\ref{conclusion}.
	
	\section{Observations and data reduction}\label{obs}
	
	We observed \object{MASCARA-1b} for two nights with the CARMENES spectrograph \citep{Quirrenbach2018} installed at the 3.5 m telescope of the Calar Alto Observatory. The spectrograph has two channels that cover 520--960\,nm in the visible (VIS) with a resolution of 94\,600 and 960--1710\,nm with a resolution of 80\,400 in the near-infrared (NIR). The first observation was performed on the night of August 4, 2020 at orbital phases before the planetary secondary eclipse (program ID: F20-3.5-019). The second observation was performed on the night of August 7, 2022 at orbital phases after the secondary eclipse as part of the CARMENES Legacy program. We observed the target with the VIS and NIR channels simultaneously. The first observing night was interrupted by a cloud passing by and was paused for about one hour. The observation of the second night was performed continuously. The detailed observing logs are presented in Table \ref{observation_logs}.
	\begin{table*}
		\large
		\renewcommand\arraystretch{1.35}
		\caption{Observation logs.}             
		\label{observation_logs}      
		\centering          
		\begin{tabular}{l c c c c c c}    
			\hline\hline
			& Date & Airmass change & Exposure time\,(s) & $N_{\mathrm{spectra}}$ & Phase coverage & $S/N$ range\tablefootmark{(a)}\\
			\hline
			VIS night-1 & 2020-08-04 & 1.67--1.12--1.13 & 120 & 66 & 0.382--0.460 & 51--85\\
			VIS night-2 & 2022-08-07 & 1.22--1.12--1.34 & 300 & 42 & 0.537--0.616 & 51--105\\
			NIR night-1 & 2020-08-04 & 1.67--1.12--1.13 & 126 & 68 & 0.382--0.460 & 27--61\\
			NIR night-2 & 2022-08-07 & 1.22--1.12--1.34 & 306 & 42 & 0.537--0.616 & 41--72\\
			\hline
			
		\end{tabular}
		\tablefoot{\tablefootmark{(a)}{The S/N per pixel was measured at $\sim 6731\,\text{\AA}$ for the VIS data and $\sim 12678\,\text{\AA}$ for the NIR data.}}
	\end{table*}
	
	The raw data were reduced using the CARMENES pipeline {\tt CARACAL} \citep{Zechmeister2014, Caballero2016}. The pipeline handles standard data reduction procedures including dark subtraction, flat fielding, wavelength calibration, and spectrum extraction. The pipeline delivers extracted one-dimensional spectra in the observer's rest frame that contain 61 spectral orders in the VIS and 28 spectral orders in the NIR. The noise value for each data point of the spectrum is also delivered by the pipeline.
	
	The original spectrum was cleaned to mitigate the influence of significant noise in certain data points. We masked the wavelength points with a low signal-to-noise ratio (S/N) for all the spectra. In the case of VIS data, wavelength points with an S/N of < 40 were masked, while for NIR data, data points with S/N < 30 on the first night and S/N < 40 on the second night -- about 20\% of the total data points -- were masked. Additionally, data points with wavelengths below 5450$\text{\AA}$ and above 8920$\text{\AA}$ were removed for the VIS data. For the NIR data, spectra from Echelle orders 45-43, which correspond to the strong water absorption band around 1.4 $\mathrm{\mu m}$, were excluded. We normalized the spectra for each order using a seventh-order polynomial fit. Subsequently, to eliminate potential outliers, such as strong sky emission lines, a $5\sigma$ clip was applied to each order as an additional step.
	
\begin{figure}
		\centering
		\includegraphics[width=9cm]{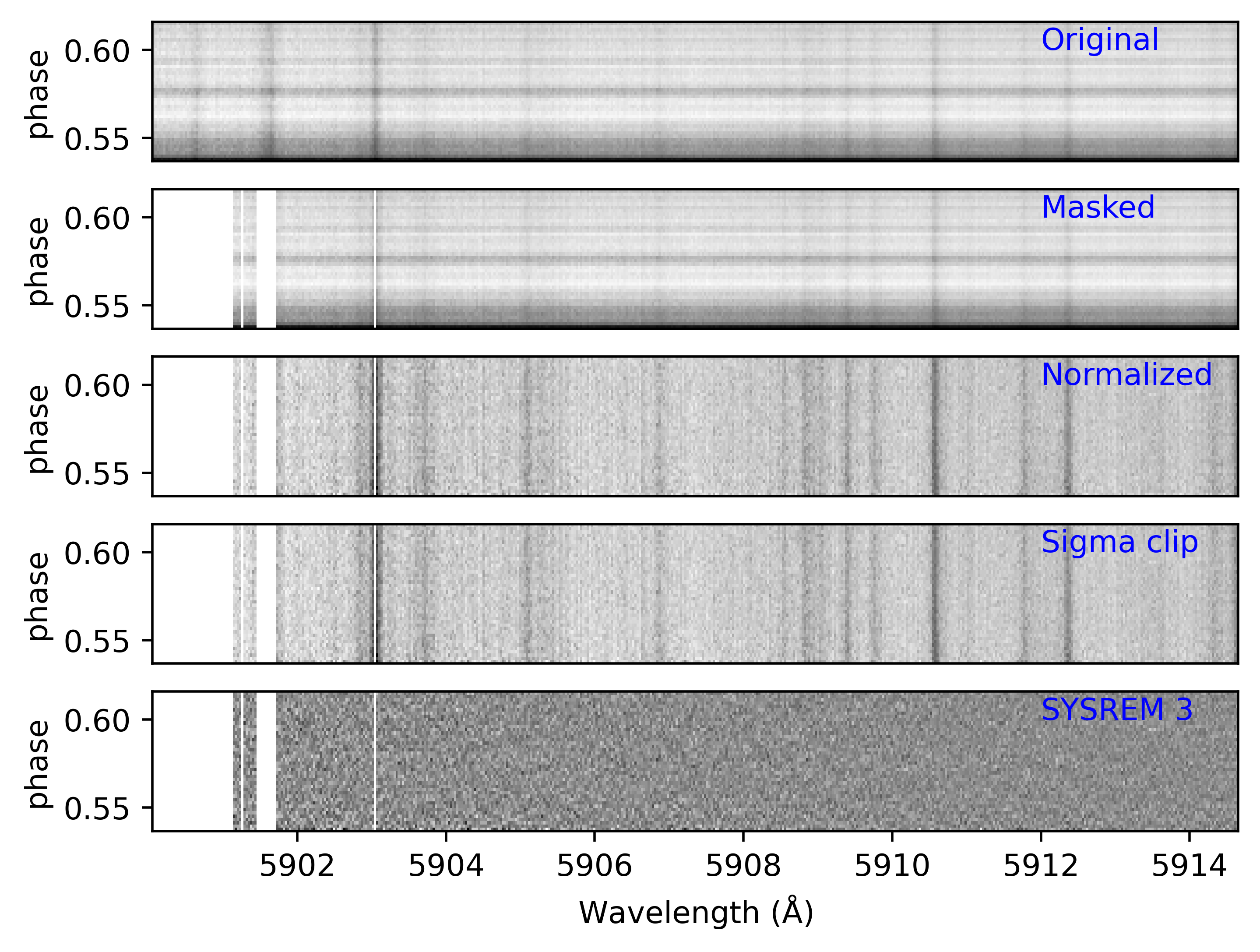}
		\caption{Example of the data reduction procedure. These are spectra from the second night. The original spectrum, masked spectrum, normalized spectrum, spectrum after the $5\sigma$ clip, and spectrum after the third \texttt{SYSREM} iteration are presented from top to bottom, respectively.}
		\label{example_spectra}
	\end{figure}

	The \texttt{SYSREM} algorithm \citep{sysrem} was employed to effectively eliminate both the telluric and stellar lines from the spectrum. The normalized spectral matrix and accompanying noise were provided as inputs to the \texttt{SYSREM} algorithm. The noise was computed via error propagation. We ran \texttt{SYSREM} ten consecutive times on the normalized data, and each time we got a \texttt{SYSREM} model. Then we divided the normalized data by the \texttt{SYSREM} model to get the residual spectral matrix that was finally used for cross-correlation. An example of the data reduction procedure is presented in Fig.\ref{example_spectra}.
	
	\section{Method}\label{method}
	
	\subsection{Model spectra}
	\label{model}
	Before performing the cross-correlation, we calculated the corresponding thermal emission spectrum model of \object{MASCARA-1b}. Our assumed atmospheric model is as described in \citet{temperature_template}. We used a two-point temperature-pressure profile, with the higher altitude point $(\textit{T}_\mathrm{1}, \textit{P}_\mathrm{1})$ set at (4500\,K, $10^{-3}$\,bar) and the lower altitude point $(\textit{T}_\mathrm{2}, \textit{P}_\mathrm{2})$ set at (2000\,K, $10^{-1.5}$\,bar). Additionally, we assumed constant mixing ratios for each species, matching the metal abundance found in the Sun.  For example, the volume mixing ratio was set as $10^{-4.59}$ for \ion{Fe}{i} and $10^{-7.14}$ for \ion{Ti}{i}. The metal opacities were derived from the Kurucz line list \citep{kurucz}; CrH opacities were obtained from the MoLLIST database \citep{CrH_opacity,MoLLIST}; OH opacities were computed based on the HITEMP line list \citep{OH_opacity};  AlO, FeH, and TiO opacities were  sourced from the ExoMol line list \citep{FeH_opacity,AlO_opacity,exomol,TiO_opacity}, and VO opacities were calculated using the Plez line list \citep{PRT}. Some other parameters are shown in Table \ref{MASCARA-1b's parameters}. Then we used \texttt{petitRADTRANS} \citep{PRT} to calculate the emission spectrum of the planet ($F_\mathrm{p}$), and we assumed that the star's emission spectrum ($F_\mathrm{s}$) is a black body radiation spectrum. Since we observed the entire star-planet system and normalized the observed spectra, the final model spectrum was thereby calculated as $(F_\mathrm{s} + F_\mathrm{p}) / F_\mathrm{s}$.
	
	\begin{table}
		\small
		\renewcommand\arraystretch{1.2}
		\caption{Parameters of MASCARA-1b.}      
		\label{MASCARA-1b's parameters}      
		\centering                         
		\begin{tabular}{l c c}        
			\hline\hline               
			Parameter  & Symbol(unit) &  Value\\   
			\hline                         \textit{The star}\\
			Effective temperature & $\textit{T}_{\mathrm{eff}}\,(\mathrm{K})$ & $7554\pm150^{(a)}$\\
			Radius & $\textit{R}_\star\,(R_\odot)$ & $2.1\pm0.2^{(a)}$\\
			Mass & $\textit{M}_\star\,(M_\odot)$ & $1.72\pm0.07^{(a)}$\\
			Systemic velocity & $\textit{v}_{\mathrm{sys}}\,(\mathrm{km\,s^{-1}})$
			& $11.20\pm0.08^{(a)}$\\
			Metallicity & [Fe/H]\,(dex) & $0.15\pm0.15^{(b)}$\\
			\hline 
			\textit{The planet}\\
			Radius & $\textit{R}_\mathrm{p}\,(R_\mathrm{J})$ & $1.5\pm0.3^{(a)}$\\
			Mass & $\textit{M}_\mathrm{p}\,(M_\mathrm{J})$ & $3.7\pm0.9^{(a)}$\\
			Surface gravity & $\mathrm{log\,\textit{g}
				\,(log\,cgs)}$ & $3.63^{+0.29}_{-0.27}$\\
			Inclination	& $\textit{i}_\mathrm{p}\,(\mathrm{degree})$ & $87^{+2}_{-3}$ \tablefootmark{(a)}\\ 
			Orbital period & $\textit{P}\,(\mathrm{day}$) & $2.1487738$\\
			~ & ~ & $~~~~~~\pm0.0000009^{(b)}$\\
			Transit epoch & $\textit{T}_{\mathrm{0}}\,(\mathrm{BJD})$ & $2458833.48815$\\
			~ & ~ & $~~~~~~\pm0.00009^{(b)}$\\
			Transit duration & $\textit{T}_{\mathrm{14}}\,(\mathrm{hours})$ &  $4.05\pm0.03^{(a)}$\\
			RV semi-amplitude & $\textit{K}_\mathrm{p}\,(\mathrm{km\,s^{-1}})$ & $197.4\pm2.7$\\
			
			\hline
		\end{tabular}
		\tablebib{(a) \citet{MASCARA_parameters}. (b) \citet{spi}.}
	\end{table}
	
	\subsection{Cross-correlation}
	We first subtracted the continuum value of unity for both the model spectrum and the residual spectrum. Then we shifted the model spectrum from $-500\,\mathrm{km\,s^{-1}}$ to $500\,\mathrm{km\,s^{-1}}$ with steps of $1\,\mathrm{km\,s^{-1}}$ to generate a grid of spectral templates. The cross-correlation function (CCF) was calculated as
	\begin{eqnarray}
		\mathrm{CCF} = \sum \frac{r_i m_i}{\sigma_i^2} ,\label{ccf}
	\end{eqnarray}
	where $r_i$ is the residual spectrum, $m_i$ is the model spectrum, and $\sigma_i$ is the noise of the observed spectrum.
	
\begin{figure}
		\centering
		\includegraphics[width=9cm]{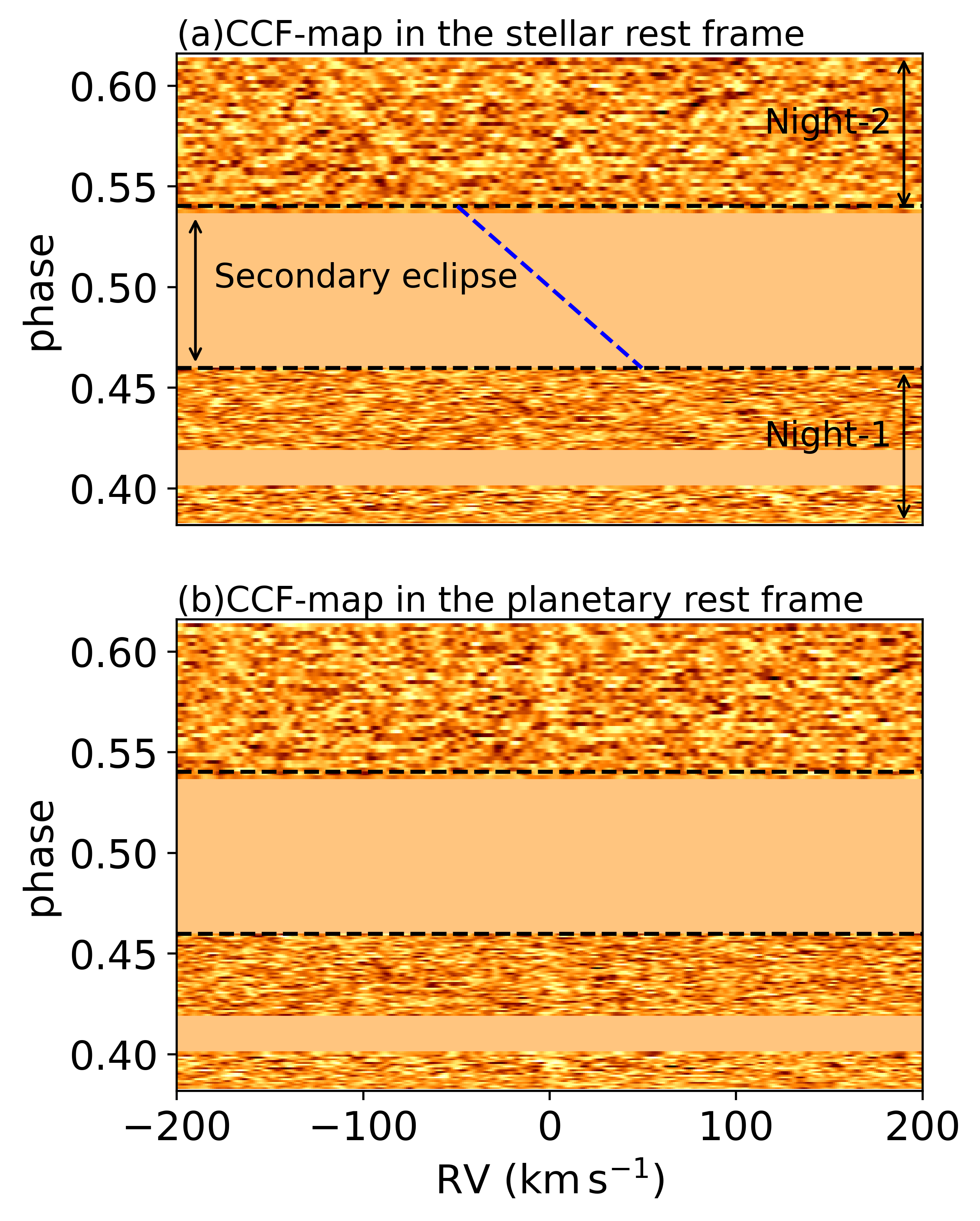}
		\caption{Cross-correlation functions for \ion{Fe}{i} from the two nights. \textit{Upper panel}: CCF map in the stellar rest frame. \textit{Lower panel}: CCF map in the planetary rest frame, which was shifted relative to the stellar rest frame by assuming a theoretical $K_\mathrm{p}$ of  $197.4\,\mathrm{km\,s^{-1}}$ obtained through Eq.(\ref{Kp_calculate}).  The horizontal dashed lines indicate the beginning and end of the secondary eclipse. The dashed blue line denotes the planetary orbital motion RV.}
		\label{ccf_map}
	\end{figure}

	The initial CCF was calculated in the observer's rest frame. We subsequently shifted each CCF into the planetary stationary frame with a planetary radial velocity (RV) of
	\begin{eqnarray}
		v_\mathrm{p} = v_{\mathrm{sys}} - v_{\mathrm{bary}} + K_\mathrm{p} \sin(2 \pi \phi) + \Delta v ,
		\label{speed}
	\end{eqnarray} 
	where $v_{\mathrm{sys}}$ is the systemic velocity (shown in Table \ref{MASCARA-1b's parameters}), $v_{\mathrm{bary}}$ is the barycentric velocity of the observer on the Earth,  $K_\mathrm{p}$ is the orbital velocity semi-amplitude, $\phi$ is the orbital phase, $\phi = 0$ corresponds to the middle transit time, and $\Delta v$ is the velocity deviation from the planetary rest frame.  The $\Delta v$ parameter accounts for the uncertainty of $v_\mathrm{sys}$, $v_\mathrm{bary}$ and $K_\mathrm{p} \sin(2 \pi \phi)$ and can also serve as an indication of planetary rotation and atmospheric circulation.  $K_\mathrm{p}$ and $\Delta v$ are the unknown quantities that we aim for. The obtained CCF map for \ion{Fe}{i} is shown in Fig.\ref{ccf_map}. We repeated this procedure for various $K_\mathrm{p}$ ranges from $80\,\mathrm{km\,s^{-1}}$ to $300\,\mathrm{km\,s^{-1}}$. For each $K_\mathrm{p}$, we summed all CCFs in the planetary frame along the time dimension to obtain a one-dimensional array. By applying the same procedure to each $K_\mathrm{p}$, we obtained the $K_\mathrm{p}$--$\Delta v$ map. To obtain an S/N map of the investigated chemical species, we normalized the $K_\mathrm{p}$--$\Delta v$ map with an estimate of the noise level, for which we used the average value in the $K_\mathrm{p}$--$\Delta v$ map, excluding the region around the central peak. The obtained S/N maps for \ion{Fe}{i} and \ion{Ti}{i} are shown in Fig.\ref{kpmap}. An evident signal at the theoretical $K_\mathrm{p}$ and $\Delta v$ = 0 confirms the existence of the chemical species.
	
	%    We repeated this procedure for various $K_\mathrm{p}$ ranges from $80\,\mathrm{km\,s^{-1}}$ to $300\,\mathrm{km\,s^{-1}}$ and summed all CCFs along the time dimension in the planetary frame for each $K_\mathrm{p}$ to get a two-dimensional array \textbf{($K_\mathrm{p}$--$\Delta v$ map)}.
	
	\section{Detection of chemical species}\label{result}
\begin{figure*}
		\centering
		\includegraphics[width=17cm]{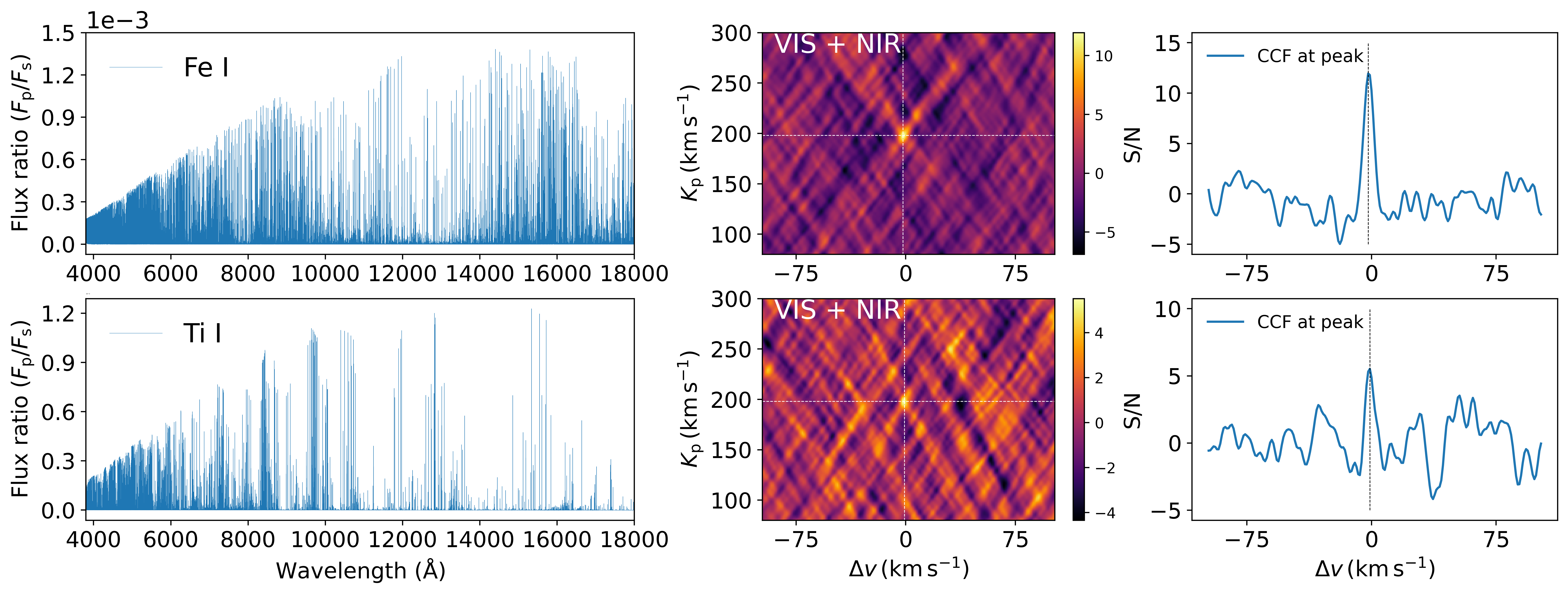}
		\caption{Model spectra and S/N maps of \ion{Fe}{i} and \ion{Ti}{i}. \textit{Left panels}: Model spectrum of each chemical species. \textit{Middle panels}: S/N map of each chemical species. The signals from both the VIS and NIR channels are combined. The dotted white line is the position of $K_\mathrm{p} - \Delta v$ where the S/N is maximum. \textit{Right panels}: CCF at the $K_\mathrm{p}$ of the maximum S/N. }
		\label{kpmap}
	\end{figure*}

	We detected signals of \ion{Fe}{i} and \ion{Ti}{i}, and the corresponding S/N map is presented in Fig.\ref{kpmap}.
	We chose the \texttt{SYSREM} iteration number that yields the highest significance value on the S/N map (see Fig.\ref{sysrem_iter_Fe} for \ion{Fe}{i} and Fig.\ref{sysrem_iter_Ti} for \ion{Ti}{i}). 
	The \ion{Fe}{i} signal is relatively strong and can be identified directly on the CCF map (Fig.\ref{ccf_map}). For demonstration purposes, we combined the VIS and NIR CCF maps, as is shown in Fig.\ref{ccf_map}. Due to the slight phase difference between the VIS and NIR data, we interpolated the VIS CCF map to a grid of phase values as the NIR data and then directly summed them.
	
\begin{figure}
		\centering
		\includegraphics[width=9cm]{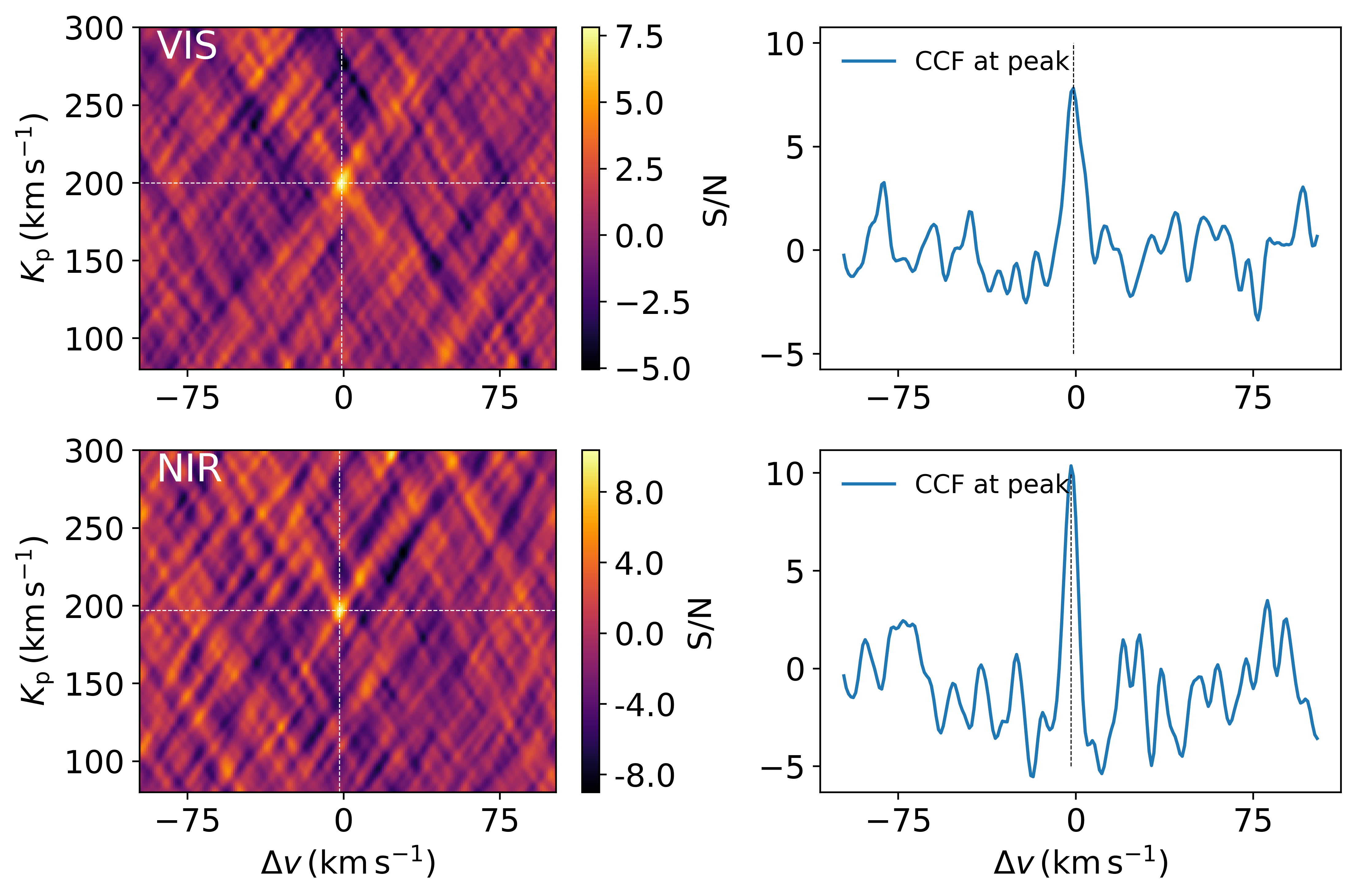}
		\caption{S/N maps of \ion{Fe}{i} from the VIS channel (upper panels) and the NIR channel (lower panels).}
		\label{kpmap_Fe}
	\end{figure}

	The theoretical $K_\mathrm{p}$ value of MASCARA-1b was calculated to be $197.4 \pm 2.7\,\mathrm{km\,s^{-1}}$ using the equation
	\begin{eqnarray}
		K_\mathrm{p} = \left(\frac{2\pi G\cdot M_\star}{P}\right)^{\frac{1}{3}}\cdot \sin i_\mathrm{p} ,
		\label{Kp_calculate}
	\end{eqnarray} 
	where G is the gravitational constant, \textit{P} is the orbital period, $M_\star$ is the stellar mass, and $i_\mathrm{p}$ is the orbital inclination. The \ion{Fe}{i} signal achieves a maximum S/N ($\sim$ 12.1) at $K_\mathrm{p} = 197.9^{+2.5}_{-2.6}\,\mathrm{km\,s^{-1}}$ and $\Delta v = -1.6 \pm 1.2\,\mathrm{km\,s^{-1}}$, aligned with the theoretical $K_\mathrm{p}$ value. \ion{Fe}{i} is detected in both VIS and NIR spectra (see Fig.\ref{kpmap_Fe}). The maximum S/N ($\sim$ 4.9) for \ion{Ti}{i} is located at $K_\mathrm{p} = 199.1^{+3.5}_{-4.2}\,\mathrm{km\,s^{-1}}$ and $\Delta v = -1.3^{+1.6}_{-1.7}\,\mathrm{km\,s^{-1}}$, which is consistent with the theoretical $K_\mathrm{p}$ value. Although the \ion{Ti}{i} signal is weaker than that of \ion{Fe}{i}, \ion{Ti}{i} is also detected in both VIS and NIR spectra.
	All of the detection results are summarized in Table \ref{detection_results}.
	%Apart from the weak signal detected in the VIS night-1 data,\ion{Ti}{i} exhibits strong signals in the data from other parts.

	\begin{table}
		\renewcommand\arraystretch{1.44}
		\caption{Summary of cross-correlation results.}      
		\label{detection_results}      
		\centering                         
		\begin{tabular}{c c c c}      
			\hline
			\hline
			chemical species & S/N &  $K_\mathrm{p}\,[\mathrm{km\,s^{-1}}]$  & $\Delta v\,[\mathrm{km\,s^{-1}}]$ \\
			\hline
			\ion{Fe}{i} & 12.05 & $197.9^{+2.5}_{-2.6}$ & $-1.6 \pm 1.2$ \\
			\ion{Ti}{i} & 4.87 & $199.1^{+3.5}_{-4.2}$ & $-1.3^{+1.6}_{-1.7}$\\
			
			\hline
		\end{tabular}
	\end{table}
	
	\setlength{\parskip}{0.2cm plus4mm minus3mm}%用于调节图片下方的空白
	
	We also applied the identical method to search for \ion{Al}{i}, AlO, \ion{Ba}{i}, \ion{Ca}{i}, \ion{Co}{i}, \ion{Cr}{i}, CrH, \ion{Fe}{ii}, FeH, \ion{Li}{i}, \ion{Na}{i}, \ion{Mg}{i}, \ion{Mn}{i}, \ion{Sc}{i}, \ion{Ti}{ii}, TiO, \ion{V}{i}, \ion{V}{ii}, VO, \ion{Si}{i}, and OH. Although some tentative signals were detected in some of the data for certain chemical species (see Fig.\ref{some_signal}) -- for example, the signal of \ion{Si}{i} was detected in the NIR data from the first night -- the corresponding cross-correlation signals were not obtained when all the observation data were combined. These tentative signals could possibly be random noise. Therefore, we do not claim to have detected any of these chemical species in MASCARA-1b with the CARMENES data. A selection of S/N maps are presented in Fig.\ref{Kpmap_no_signal}, Fig.\ref{Kpmap_no_signal2}, Fig.\ref{Kpmap_no_signal3}, and Fig.\ref{Kpmap_no_signal4}.
	
	The CARMENES detections of \ion{Fe}{i} and \ion{Ti}{i} are consistent with the previous PEPSI detection by \citet{MASCARA-1b_CrI}; however, we were not able to detect \ion{Cr}{i}, which was reported in their work.  This may be due to the difference in the wavelength coverage of the two instruments. Compared to PEPSI's wavelength domain ($4800-5441\,\text{\AA}$ and $7419-9067\,\text{\AA}$), the wavelength coverage of CARMENES ($5200-17100\,\text{\AA}$) does not contain a significant number \ion{Cr}{i} spectral lines, especially after we masked the points with wavelengths below $5450\,\text{\AA}$ (see Fig.\ref{Kpmap_no_signal}).
	
	The presence of \ion{Ti}{i} and \ion{V}{i} in the atmosphere of UHJs is of great interest because of the importance of their oxides (i.e., TiO and VO; \citealt{inversions_1, inversions_2}).
	The strong absorption of stellar radiation by TiO and VO has long been believed to be a possible cause of the formation of the temperature inversion layer in the atmosphere of UHJs. However, TiO and VO molecules are not commonly detected in UHJs. Therefore, it is proposed that Ti and V can be cold-trapped on the nightside where the temperature is lower than the condensation temperature of \ion{Ti}{i} and \ion{V}{i}, which leads to the absence of TiO and VO features in the transmission and emission spectra.
	In recent years, \ion{Ti}{i} depletion phenomena have been discovered on WASP-121b \citep{WASP-121b-cold_trap} and WASP-76b \citep{WASP-76b-cold_trap}. However, we detected \ion{Ti}{i} on MASCARA-1b, indicating that there may not be a Ti cold-trap on MASCARA-1b. 
	This may be due to the high equilibrium temperature of MASCARA-1b ($T_\mathrm{eq} = 2594.3^{+1.6}_{-1.5}\,\mathrm{K}$).
	
	%It is believed that this may be due to the night-side temperature of the planet being lower than the condensation temperature of \ion{Ti}{i}, causing \ion{Ti}{i} to condense on the night-side and be cold-trapped (some believe \ion{Ti}{i} condenses into \ch{CaTiO3} (perovskite)), preventing \ion{Ti}{i} from cycling back to the detection area of emission and transmission spectra. 
	
	\section{Retrieval of atmospheric properties}\label{retrieval}
	
	\subsection{Retrieval method}
	Atmospheric retrieval techniques for high-resolution spectroscopy were motivated by \citet{retrieval_methods_1}, \citet{retrieval_methods_2}, \citet{retrieval_method_3} and  \citet{temperature_template} in recent years. We retrieved the atmospheric properties with the observed emission spectra of \ion{Fe}{i} and \ion{Ti}{i}, employing the latest retrieval method described in \cite{new_retrieval_method}.	
	
	Forward models were calculated using \texttt{petitRADTRANS}, assuming a two-point \textit{T}-\textit{P} profile for the atmospheric structure. We set the high-altitude temperature, $T_1$, and the low-altitude temperature, $T_2$, as free parameters. The range of $T_1$ is from 2000 to 6000$\,$K and the range of $T_2$ is from 1000 to 5000$\,$K. We defined the pressure point, $\mathrm{log}\,P_\mathrm{1}$, at high altitude as a free parameter, ranging from -7 to 0 (log bar). Simultaneously, the pressure difference, $dP$, between high and low altitude points was utilized to indirectly represent the pressure at low altitude ($\mathrm{log}\,P_\mathrm{2}$ = $\mathrm{log}\,P_\mathrm{1} + dP$). The value of $dP$ is also a free parameter, varying between 0 and +7. Since $dP$ was set as > 0, we ensured that $P_\mathrm{2}$ is always at a higher pressure than $P_\mathrm{1}$.
	We set the atmosphere structure in \texttt{petitRADTRANS} as 25 pressure layers, evenly spaced in log(\textit{P}) from 1\,bar to $10^{-8}$\,bar. Such a setup means that the lower boundary of the atmosphere is set at 1\,bar.
	Then we generated a model spectral matrix with the same data structure as the observed residual spectral matrix. The model spectra were translated into the observer's stationary frame using the RV value described in Eq.(\ref{speed}).	
	
	Regarding the treatment of the mixing ratio of the chemical species, we employed two different methods. The first method was free retrieval, in which the volume mixing ratio of each chemical species is assumed to be constant across all pressure levels of the planetary atmosphere. For example, we denoted the logarithm of the \ion{Fe}{i} volume mixing ratio as log (Fe). The second method is chemical equilibrium retrieval. We assumed the elemental abundance of each atom or ion to be a free parameter. For example, the iron elemental abundance ([Fe/H]) was assumed to be a free parameter in the retrieval. Here, [Fe/H] refers to the logarithm of the ratio between the Fe/H of the planet and the Fe/H of the Sun. The actual profile of the \ion{Fe}{i} mixing ratio was then computed under the chemical equilibrium assumption for given [Fe/H] values. We used the chemical equilibrium module \texttt{easyCHEM} of \texttt{petitCode} \citep{petitCode2015,petitCode2017}, which consists of 84 chemical species, including \ion{Fe}{i}, \ion{Fe}{ii}, \ion{Ti}{i}, \ion{Ti}{ii}, \ion{V}{i}, \ion{V}{ii}, TiO, VO, and FeH.
	
	In the retrieval process, line profile broadening was also taken into consideration. Pressure and thermal broadening were already accounted during the calculation of the petitRADTRANS opacity grid. Additionally, the model spectrum was convolved with the instrumental profile. Gaussian profiles were assumed for the instrumental broadening, with a VIS spectral resolution of \textit{R} = 94\,600 and a NIR spectral resolution of \textit{R} = 80\,400. The \texttt{broadGaussFast} code from \texttt{PyAstronomy} \citep{pyastrnomy} was used for this convolution.  Moreover, the influence of planetary rotation on spectral line broadening was considered. We used the rotational profile in \citet{v_eq_spectra},
	\begin{eqnarray}
		G(x) = \frac{2(1-\varepsilon)(1-x^2)^{\frac{1}{2}} + \frac{\pi\varepsilon}{2}(1-x^2)}{\pi(1-\frac{\varepsilon}{3})}
		,\end{eqnarray} 
	to calculate the rotational broadening, where $x = \ln(\lambda/\lambda_0)\cdot c/(v_\mathrm{eq}\sin i)$, $\lambda_0$ is the central wavelength of the spectral line, $\varepsilon$ is the limb darkening coefficient, and $v_\mathrm{eq}$ is the equatorial rotation velocity. We assumed a linear limb darkening model and set $\varepsilon$ to 1, implying that the radiation from the limb region of the planet does not contribute to the total radiation flux. Since the atmospheric distribution of planets is inhomogeneous, $\varepsilon$ represents both the limb darkening and the degree of inhomogeneity of the thermal distribution. Finally, we convolved the model spectrum with this rotational profile in velocity space (i.e., $\ln\lambda$).
	
	During the data reduction process, the observed spectrum underwent the \texttt{SYSREM} algorithm, altering the intensity and profile of the planetary lines. The actual distortion of the line profile is also phase-dependent. Therefore, it is theoretically necessary to perform the same \texttt{SYSREM} processing on the model matrix before fitting it with the residual spectral matrix. Nevertheless, implementing the \texttt{SYSREM} algorithm for each template spectral matrix would be time-consuming. We utilized the fast \texttt{SYSREM} filtering technique described by \cite{fast_sysrem} to mitigate this issue. We generated a filter matrix while performing \texttt{SYSREM} on the observed spectral matrix. The filter matrix was generated during the \texttt{SYSREM} process on the observed spectral matrix. The filter matrix was then multiplied with the model spectrum matrix, resulting in the final model spectral matrix. Furthermore, a Gaussian high-pass filter with a Gaussian $\sigma$ of 31 points was applied to both the final residual matrix and model spectral matrix to remove any remaining broadband features.
	
	We performed the atmospheric retrieval by evaluating the likelihood function with the Markov chain Monte Carlo (MCMC) simulation tool \texttt{emcee} \citep{emcee}. The logarithm likelihood function is described as 
	\begin{eqnarray}
		\ln (L) = - \sum_i \left[  \frac{(R_i - \alpha M_i)^2}{(\beta \sigma_i)^2} + \ln (2 \pi (\beta \sigma_i)^2) \right],
	\end{eqnarray}
	where $R_i$ is the residual spectral matrix, $M_i$ is the model spectral matrix, $\sigma_i$ is the residual noise matrix, $\alpha$ is the scale coefficient of the model spectrum, and $\beta$ is the scale factor of the noise. The $\alpha$ parameter accounts for the uncertainties of the systemic parameters such as $(R_\mathrm{p}/R_\mathrm{s})^2$. It also accounts for the fact that only part of the dayside hemisphere faces toward the observer at orbital phases other than 0.5. We used the residual spectral matrix corresponding to the best \texttt{SYSREM} iterations from the \ion{Fe}{i} detection. We ran the MCMC simulation with 10000 steps and 100 walkers for each free parameter. For the final retrieval results, we burned in initial 2000 steps. 
	%The initial 2000 steps in the retrieval process are burn-in, intended to eliminate any non-convergent data that may exist.

	\subsection{Retrieval results}
	% Two assumptions were made in this retrieval regarding the volume mixing ratio of \ion{Fe}{i}. It is either assumed to be free retrieval, which means it is constant throughout the atmospheric structure, or determined by using the chemical equilibrium module \texttt{easyCHEM} of \texttt{petitCode} \citep{petitCode2015,petitCode2017}. The latter assumption is based on the variation of the volume mixing ratio of all metals with the overall metallicity. Additionally, the element abundance was set to solar.
	\subsubsection{Temperature structure}
\begin{figure}
		\centering
		\includegraphics[width=9cm]{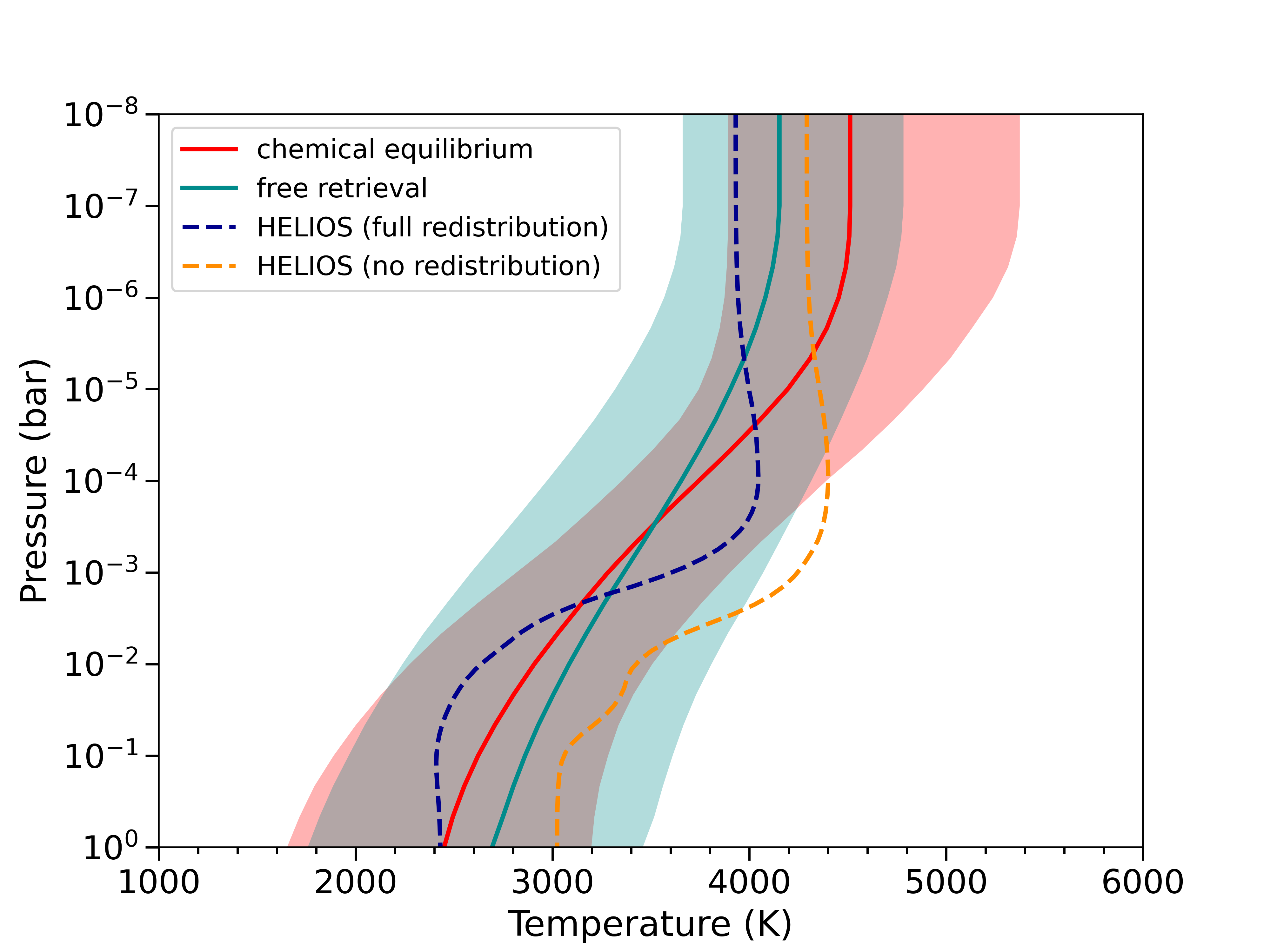} 
		\caption{Retrieved \textit{T}-\textit{P} profile and comparison with the results from the self-consistent \texttt{HELIOS} model. The \texttt{HELIOS} model was calculated assuming solar abundances for both no and full heat redistribution from dayside to nightside. The shadows are the $1\sigma$ range of the retrieved results. }
		\label{T-P_profile}
	\end{figure}

	The retrieval was performed independently for the free retrieval and chemical equilibrium retrieval. Figure \ref{T-P_profile} illustrates the final retrieved \textit{T}-\textit{P} profiles. The retrieved parameter values are summarized in Table \ref{retrieved_paremater} with posterior distributions plotted in Figs.\ref{corner_map} and \ref{corner_map2}. The retrieval results indicate the presence of a strong atmospheric inversion layer with a temperature difference of $\sim$ 1700\,K under the free retrieval. We also conducted a retrieval with $\alpha$ fixed as 1, but there was no significant difference compared to the retrieval results with free $\alpha$. Additionally, we computed self-consistent models employing the modified \texttt{HELIOS} code developed by \citet{helios}, which incorporates opacities caused by both neutral and singly ionized species, as was proposed by \citet{NLTE}. A detailed description of the \texttt{HELIOS} model calculation can be found in \citet{KELT20b_detection}. The \textit{T}-\textit{P} profiles from the \texttt{HELIOS} model are presented in Fig.~\ref{T-P_profile} for comparison. The retrieved temperature difference in the inversion layer is similar to the prediction from the \texttt{HELIOS} model. However, the location of the inversion layer (i.e., $P_1$ and $P_2$) has a large dynamic range, which is a result of the degeneracy between the location of the inversion layer and the metallicity.
	
	\begin{table*}
		\large
		\renewcommand\arraystretch{1.35}
		\caption{Parameter values from the \textit{T}-\textit{P} profile retrieval.}             
		\label{retrieved_paremater}      
		\centering
		
		\begin{tabular}{c c c c c c c c}    
			\hline\hline
			Parameter (Unit) &  free retrieval & chemical equilibrium & Boundaries\\
			\hline
			$T_\mathrm{1}$ (K) & $4200^{+600}_{-500}$ & $4500^{+900}_{-600}$ & 2000 to 6000\\
			$\mathrm{log}\,P_\mathrm{1}$ (log\,bar) & $-5.3^{+2.3}_{-1.2}$ & $-5.4^{+1.3}_{-1.0}$ & -7 to 0 \\
			$T_\mathrm{2}$ (K) & $2500^{+800}_{-1000}$ & $2300^{+800}_{-900}$ & 1000 to 5000 \\
			dP (log\,bar) & $4.0^{+2.0}_{-2.2}$ & $4.2^{+1.7}_{-1.8}$ & 0 to +7 \\
			$\mathrm{log}\,P_\mathrm{2}$ (log\,bar) & $-1.0^{+2.3}_{-2.2}$ & $-1.2^{+1.6}_{-1.4}$ & ...\\
			$\beta$ & $0.90\pm{0.0003}$ & $0.90\pm{0.00025}$ & 0 to 10\\
			$\alpha$ & $1.5^{+1.6}_{-0.8}$ & $1.5^{+1.4}_{-0.6}$ & 0 to 10\\
			log\,(Fe) & $-3.6^{+1.6}_{-1.8}$ & ... & -10 to 0\\{}
			[Fe/H] (dex) & $\sim 1.0^{+1.6}_{-1.8}$ \tablefootmark{(a)} & $1.4^{+1.1}_{-1.3}$ & -3 to +3 \\
			log\,(Ti) & $-7.1^{+1.8}_{-1.7}$ & ... & -10 to 0\\{}
			[Ti/H] (dex) & $\sim 0.0^{+1.8}_{-1.7}$ \tablefootmark{(a)} & $0.8^{+1.2}_{-1.5}$ & -3 to +3 \\
			log\,(V) & $< -7.7$ & ... & -10 to 0\\{}
			$v_\mathrm{eq}$ ($\mathrm{km\,s^{-1}}$) & $4.4^{+1.6}_{-2.0}$ & $4.1^{+1.6}_{-1.9}$ & 0 to 20 \\
			$\mathrm{\Delta} v$ ($\mathrm{km\,s^{-1}}$) & $-1.2^{+0.7}_{-0.6}$ & $-1.2\pm{0.5}$ & -20 to 20 \\
			$K_\mathrm{p}$ ($\mathrm{km\,s^{-1}}$) & $199.2^{+1.4}_{-1.5}$ & $199.4\pm1.2$ & 180 to 220 \\
			\hline
		\end{tabular}
		
		\tablefoot{\tablefoottext{a}{In free retrieval, [Fe/H] is approximately calculated by assuming that the Fe element exists mainly in the form of \ion{Fe}{i} and the atmosphere is mainly composed of \ion{H}{i}.}}
	\end{table*}
	
\begin{figure}
		\centering
		\includegraphics[width=9cm]{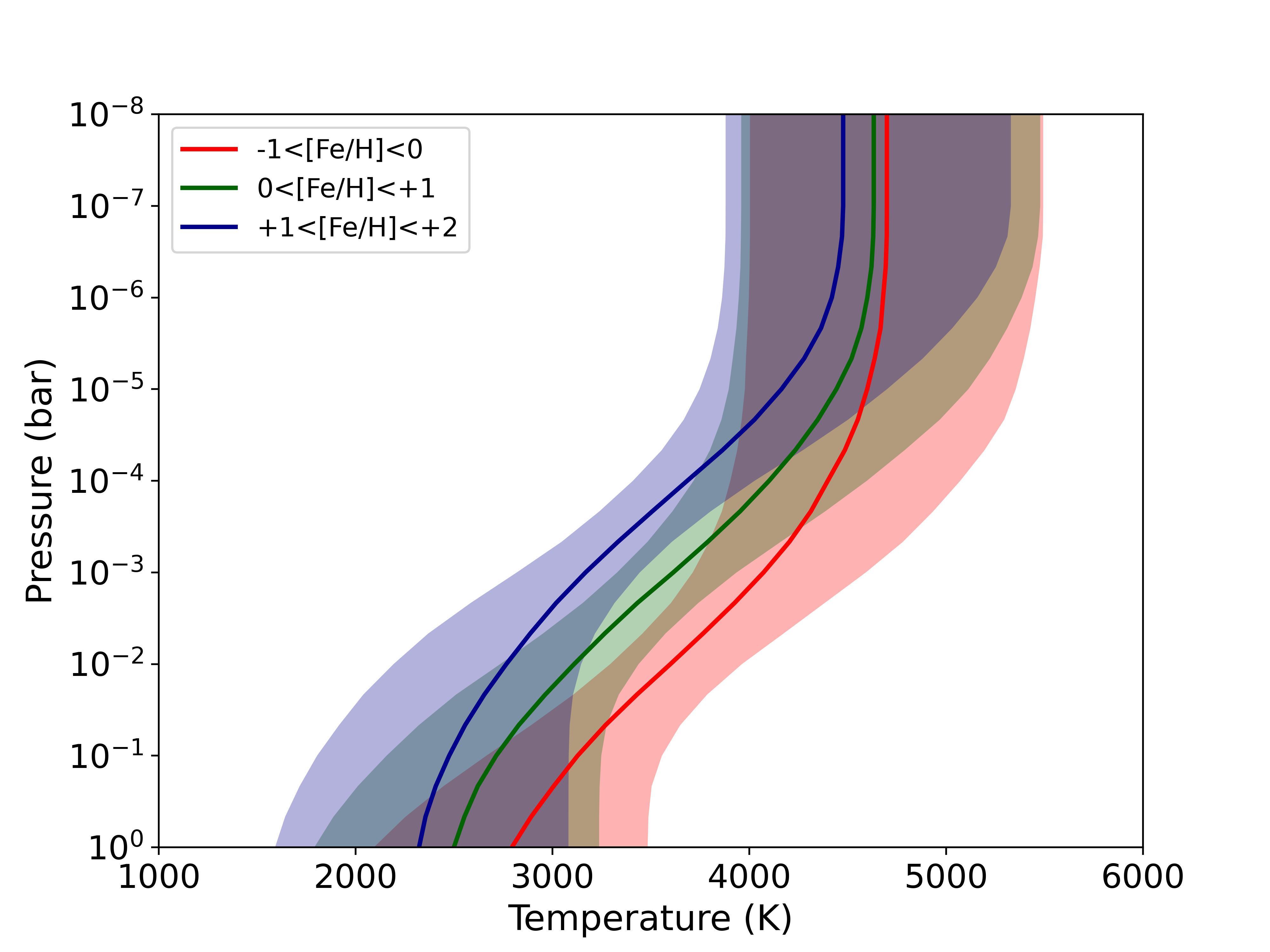} 
		\caption{Retrieved \textit{T}-\textit{P} profile for different [Fe/H] ranges.  The shadows are the $1\sigma$ range of the retrieved results. }
		\label{T-P_profile_feh}
	\end{figure}

	Compared with the retrieved results using CRIRES$^+$ data in \cite{CRIRES+_MASACRA1b}, we detected a higher altitude of the inversion layer in the atmospheric structure. This discrepancy can be attributed to several factors. First, there exists a certain degeneracy between [Fe/H] and pressure, which has been reported in previous studies (e.g.,\citealp{temperature_template}). In the retrieval process of \citet{CRIRES+_MASACRA1b}, there is an excessive constraint on the metallicity [M/H] boundary, hindering its attainment of the optimal fitting value. As a result, the position of the inversion layer can be modified by the strong boundary condition of metallicity. As is shown in Fig.\ref{T-P_profile_feh}, if we artificially restrict the range of [Fe/H] to smaller interval, the altitude of the inversion layer will correspondingly change also. When the metallicity is limited to -1<[Fe/H]<0, the pressure at the low-altitude point, $\mathrm{log}\,P_\mathrm{2}$, is $-0.3^{+1.6}_{-1.0}$. As the metallicity increases to 0<[Fe/H]<+1, $\mathrm{log}\,P_\mathrm{2}$ shifts to $-0.9^{+1.3}_{-1.1}$. For +1<[Fe/H]<+2, the pressure reduces to $\mathrm{log}\,P_\mathrm{2} = -1.6^{+1.6}_{-1.2}$. Second, the atmospheric structure is predominantly constrained by CO and \ch{H2O} lines in \citet{CRIRES+_MASACRA1b}, while our retrieval is mostly driven by \ion{Fe}{i} lines. 
	
	\subsubsection{Chemical abundance}
\begin{figure}
		\centering
		\includegraphics[width=9cm]{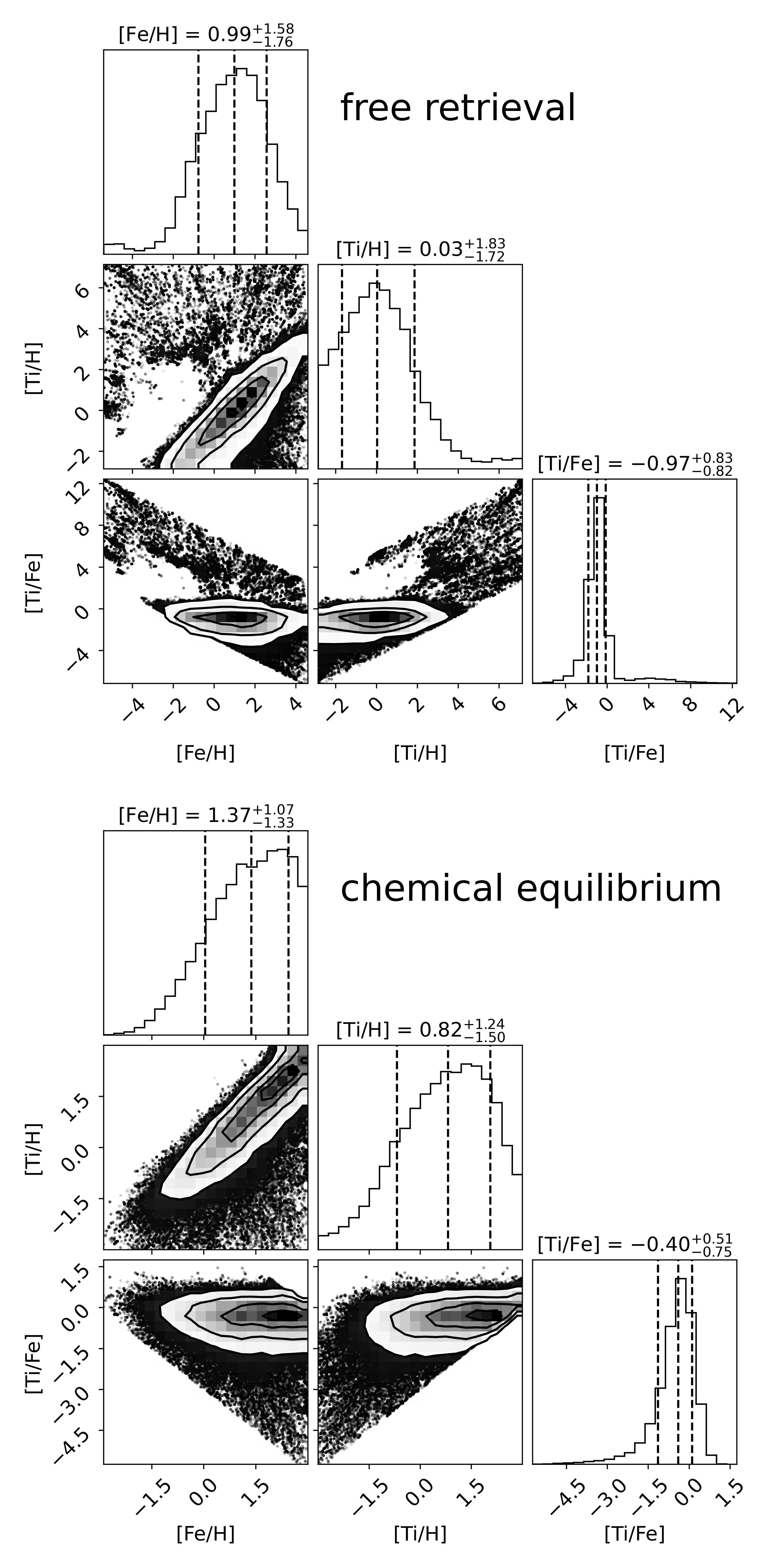} 
		\caption{Element abundances of Fe and Ti relative to the Sun. The upper panel is the result of free retrieval and the lower panel is from chemical equilibrium retrieval.}
		\label{corner_map_VMR}
	\end{figure}

	The results of the free retrieval indicate that the volume mixing ratios of \ion{Fe}{i} (${\mathrm{log\,(Fe)}}$) and \ion{Ti}{i} (${\mathrm{log\,(Ti)}}$) are $-3.6^{+1.6}_{-1.8}$ and $-7.1^{+1.8}_{-1.7}$, respectively. Since the actual ${\mathrm{log\,(Fe)}}$ and ${\mathrm{log\,(Ti)}}$ will decrease with altitude due to thermal ionization, the retrieved ${\mathrm{log\,(Fe)}}$ and ${\mathrm{log\,(Ti)}}$ should be considered as average values in the inversion layer. We further computed the elemental abundance of Fe and Ti relative to the Sun using the free retrieval results (see Fig.\ref{corner_map_VMR}). Here, we assumed that all the Fe and Ti chemical species are in the format of atomic iron and atomic titanium and that the atmosphere is mainly composed of \ion{H}{i}.
	The obtained [Fe/H] has a value $\sim 1.0^{+1.6}_{-1.8}$ and [Ti/H] has a value  $\sim 0.0^{+1.8}_{-1.7}$, which are both consistent with the solar metallicity within one $\mathrm{\sigma}$. We also calculated the relative abundance between Ti and Fe ([Ti/Fe]) using the MCMC samples and obtained [Ti/Fe] as $-1.0 \pm 0.8$.
	%Additionally, we observed a certain level of degeneracy between ${\mathrm{log\,(Fe)}}$ or ${\mathrm{log\,(Ti)}}$ and pressure ($P_\mathrm{1}$ and $P_\mathrm{2}$), which has been reported in previous studies (e.g.,\citealp{temperature_template}).
	While \ion{V}{i} was not detected, we obtained its one $\mathrm{\sigma}$ upper limit as $\mathrm{log\,(V)} < -7.7$. This upper limit is close to the solar vanadium mixing ratio ($\mathrm{log\,(V)}=-8.04$), indicating that the non-detection of \ion{V}{i} will be attributed to the detection limitation of our data rather than the depletion of vanadium in the atmosphere.
	%The theoretical volume mixing ratio of \ion{Fe}{i} in the sun is approximately -4.59, while the volume mixing ratio of \ion{Fe}{i} obtained from the retrieval results is slightly higher. However, given that the difference is only approximately 1σ1\sigma, it is considered to be within the margin of error.
	
\begin{figure}
		\centering
		\includegraphics[width=9cm]{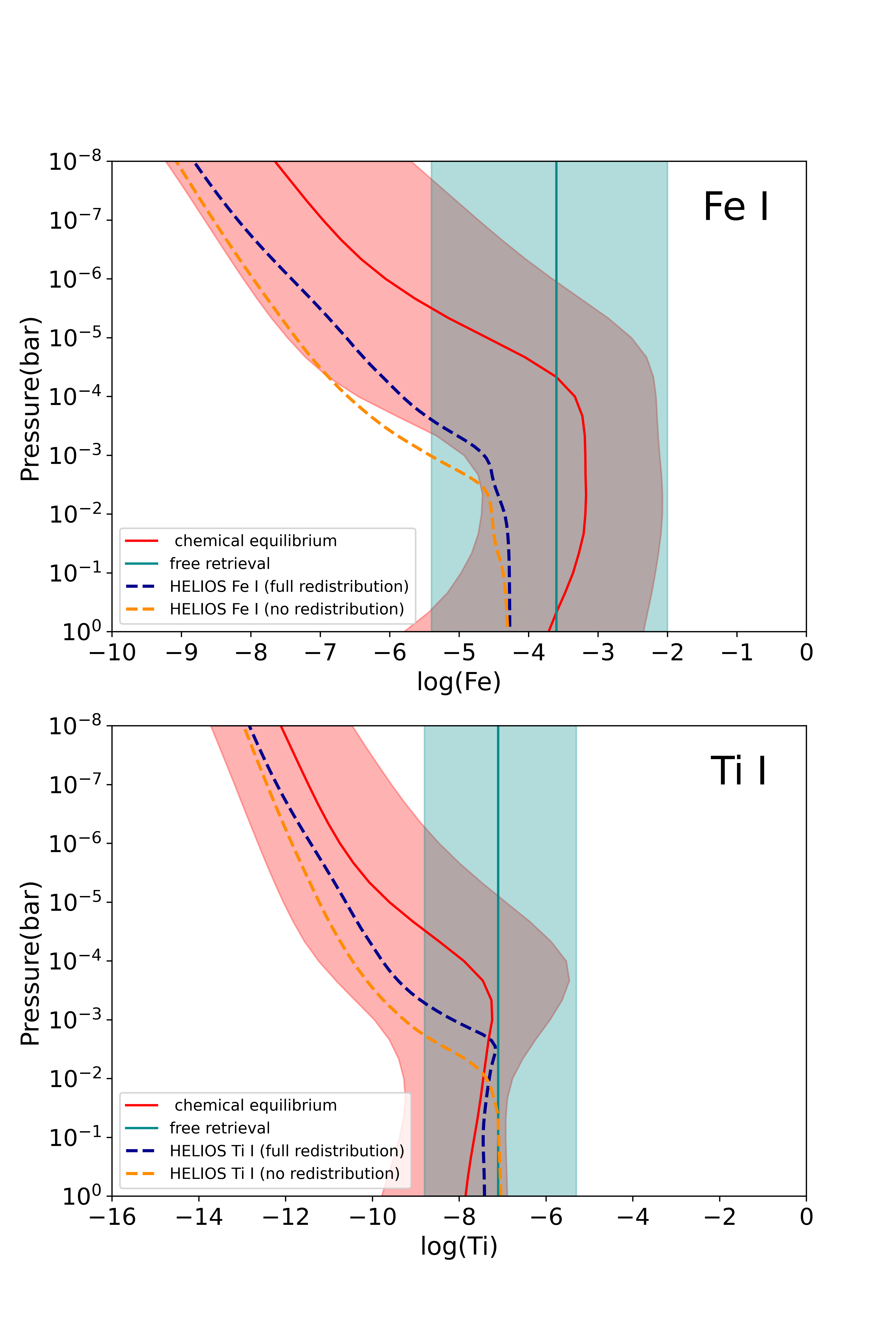}
		\caption{Retrieved volume mixing ratio of \ion{Fe}{i} (\textit{upper panel}), \ion{Ti}{i} (\textit{lower panel}), and a comparison with the results from the self-consistent \texttt{HELIOS} model. The shaded regions show the $1\sigma$ range of the retrieved results.}
		\label{VMR}
	\end{figure}

	For the chemical equilibrium retrieval, we obtained the [Fe/H] value as $1.4^{+1.1}_{-1.3}$ and the [Ti/H] value as $0.8^{+1.2}_{-1.5}$, which are both consistent with the abundances from the free retrieval. The retrieved relative abundance of [Ti/Fe] is $-0.4^{+0.5}_{-0.8}$.
	The actual mixing ratios of \ion{Fe}{i} and \ion{Ti}{i} atoms from the chemical equilibrium retrieval are plotted in Fig.~\ref{VMR}. At high altitudes, the chemical equilibrium mixing ratios are significantly smaller than the constant mixing ratios from the free retrieval, which is caused by the high thermal ionization rate at low pressures; for example, \ion{Fe}{i} dissociates into \ion{Fe}{ii} in large quantities. At very low altitudes, the chemical equilibrium mixing ratios are also slightly weaker than the free retrieval values. This is because \ion{Fe}{i} is partially condensed under high pressures, while the Ti element is mostly in the form of TiO instead of the \ion{Ti}{i} atom at very low altitudes.

	%We observed a slight decrease in the \ion{Fe}{i} abundance in the lower atmosphere within the chemical equilibrium model. This decrease can be attributed to the setting of a lower temperature limit at 1000\,K for low-altitude points, which leads to a minor condensation of \ion{Fe}{i}. The retrieved results from the chemical equilibrium model indicate [Ti/H]=1.0+1.1−1.31.0^{+1.1}_{-1.3}, however, the abundance of \ion{Ti}{i} does not appear to significantly surpass the solar abundance.  This is because some of the Ti elements exist in the form of TiO in the atmosphere.
	
	% However, a significant discrepancy exists between the results obtained from the free retrieval and those obtained from chemical equilibrium, as depicted in Fig.???\ref{VMR}. The results obtained from retrieving chemical equilibrium yield inaccurate findings, suggesting a metal abundance in the atmosphere nearly 1000 times that of the solar abundance. In this study, we propose two possible explanations. 
	Although the chemical equilibrium model includes chemical processes, compared to the free retrieval it also has certain limitations. First, nonlocal thermodynamic equilibrium effects may have some impact on the retrieval results. The presence of the inversion layer leads to higher temperatures in the planet's upper atmosphere. This causes certain energy levels of species such as \ion{Fe}{i} in the upper atmosphere to be overpopulated, emitting stronger spectral lines. This can lead to an overestimation of the mixing ratio under the local thermodynamic equilibrium assumption. Second, vertical wind may exist between the lower and upper layers of the atmosphere \citep{vertical_wind}. These vertical winds can be described as outward-expanding winds, which transport substances such as \ion{Fe}{i} from the lower atmosphere to the upper atmosphere, forming a vertical material transport in the atmosphere. According to the theory proposed by \citet{quenching}, if the timescale of transport is shorter than the chemical kinetic timescale required to maintain equilibrium between this component and other species, a component like \ion{Fe}{i} may fail to maintain a state of chemical equilibrium. This phenomenon, known as transport-induced quenching, leads to a constant volume mixing ratio for the species instead of the mixing ratio profile predicted by the chemical equilibrium model.
	% In summary, we believe that these findings suggest that the assumption of a fixed volume mixing ratio of \ion{Fe}{i} may be more applicable to the atmospheric structure of this planet than the assumption of chemical equilibrium.
	
	\subsubsection{Velocity signatures}                           
	The retrieved equatorial rotation speed, $v_\mathrm{eq}$, is $4.4^{+1.6}_{-2.0} \mathrm{km\,s^{-1}}$ in the free retrieval. If we assume that the planet is tidally locked, the corresponding equatorial rotation speed is $3.55\,\mathrm{km\,s^{-1}}$. Therefore, the retrieved $v_\mathrm{eq}$ is in agreement with the tidally locked rotation speed, considering the uncertainty range of the retrieved value. In addition to planetary rotation, atmospheric circulation such as super-rotation also contributes to the line profile; for example, a prominent super-rotation will result in a large $v_\mathrm{eq}$. Therefore, our retrieved $v_\mathrm{eq}$ hints that the planet may not have a significant equatorial super-rotation jet. 
	
	In all of the retrievals, $K_\mathrm{p}$ and $\Delta v$ were treated as free parameters. The retrieval results indicate that $K_\mathrm{p}$ is approximately $199\,\mathrm{km\,s^{-1}}$ and $\Delta v$ is approximately $-1\,\mathrm{km\,s^{-1}}$, which corroborates the findings from the cross-correlation results. The $\Delta v$ of the \ion{Fe}{i} signal deviates very slightly from the rest frame of the planet. The blueshift or redshift of the detection signal is normally an indication of planetary rotation and atmospheric circulation. However, there are several other factors that can cause the change in $\Delta v$. First, the uncertainty of the transit epoch can lead to the change in $\Delta v$. For example, if we were to use an older transit epoch ($T_0$) from \cite{MASCARA_parameters}, it would result in a deviation of approximately $-2\,\mathrm{km\,s^{-1}}$ compared to the RV calculated using the latest measured $T_0$ from \cite{spi}. Second, \citet{MASCARA_parameters} found that the measured stellar systemic velocity varies with different instruments; for example, $v_{\mathrm{sys}} = 11.20\pm 0.08\,\mathrm{km\,s^{-1}}$ for HERMES and $v_{\mathrm{sys}} = 8.50\pm 0.02\,\mathrm{km\,s^{-1}}$ for SONG. \citet{espresso} reports a systemic velocity of $9.3\pm 2.3\,\mathrm{km\,s^{-1}}$ measured using ESPRESSO data. This suggests the presence of high uncertainty in the determination of systemic velocity for early-type stars such as MASCARA-1, which in turn leads to high systematic uncertainty of the obtained $\Delta v$.
	
	\section{Conclusions}\label{conclusion}
	We observed the emission spectrum of \object{MASCARA-1b} before and after the secondary eclipse using the CARMENES spectrograph. By employing the cross-correlation technique, we detected strong signals of \ion{Fe}{i} and \ion{Ti}{i} in the planetary dayside atmosphere. The detected \ion{Fe}{i} and \ion{Ti}{i} lines are emission lines, indicating the presence of an inversion layer in the atmosphere. We also used the same method to search for other chemical species, including TiO and VO, but failed to detect them. 
	
	We performed atmospheric retrieval on the observed \ion{Fe}{i} and \ion{Ti}{i} lines. We employed two different methods to calculate the mixing ratios of the chemical substances: free retrieval and chemical equilibrium retrieval. The retrieval results indicate the existence of a strong inversion layer in the atmosphere of \object{MASCARA-1b}. 
	The retrieved elemental abundances of Fe and Ti are broadly consistent with the solar values. The obtained relative abundance of [Ti/Fe] is $-1.0 \pm 0.8$ in the free retrieval and $-0.4^{+0.5}_{-0.8}$ in the chemical equilibrium retrieval, indicating that there is no significant depletion of Ti on this planet.
	%Additionally, they suggest that this exoplanet may possess a higher iron abundance, resulting in the upward displacement of the inversion layer. The abundance of \ion{Ti}{i} is also similar to that of the Sun, indicating that there is no significant depletion of Ti on this planet. 
	We also considered rotational broadening and obtained an equatorial rotational velocity, $v_\mathrm{eq}$, of around $4.4\,\mathrm{km\,s^{-1}}$, which agrees with the tidally locked rotation velocity. During the retrieval process, we treated $K_\mathrm{p}$ and $\Delta v$ as free parameters, and the results were consistent with those obtained from cross-correlation.
	
	%     High-resolution emission spectroscopy of UHJs can effectively detect the chemical substances in their dayside atmospheres. Additionally, by employing retrieval methods, we can obtain corresponding atmospheric structures and elemental abundances. Studying exoplanet elemental abundance has great potential in enhancing our understanding of planetary formation and evolution processes.
	
	% We identified two possible explanations for the inconsistency in the retrieved \ion{Fe}{i} and \ion{Ti}{i} volume mixing ratio: NLTE effects and transport-induced quenching. 
	\begin{acknowledgements}
		We acknowledge the support by the National Natural Science Foundation of China (grant No. 42375118). 
		This publication was based on observations collected under the CARMENES Legacy+ project. 
		CARMENES is an instrument for the Centro Astron\'omico Hispano-Alem\'an (CAHA) at Calar Alto (Almer\'{\i}a, Spain), operated jointly by the Junta de Andaluc\'ia and the Instituto de Astrof\'isica de Andaluc\'ia (CSIC).
		
		The authors wish to express their sincere thanks to all members of the Calar Alto staff for their expert support of the instrument and telescope operation.
		
		CARMENES was funded by the Max-Planck-Gesellschaft (MPG), 
		the Consejo Superior de Investigaciones Cient\'{\i}ficas (CSIC),
		the Ministerio de Econom\'ia y Competitividad (MINECO) and the European Regional Development Fund (ERDF) through projects FICTS-2011-02, ICTS-2017-07-CAHA-4, and CAHA16-CE-3978, 
		and the members of the CARMENES Consortium 
		(Max-Planck-Institut f\"ur Astronomie,
		Instituto de Astrof\'{\i}sica de Andaluc\'{\i}a,
		Landessternwarte K\"onigstuhl,
		Institut de Ci\`encies de l'Espai,
		Institut f\"ur Astrophysik G\"ottingen,
		Universidad Complutense de Madrid,
		Th\"uringer Landessternwarte Tautenburg,
		Instituto de Astrof\'{\i}sica de Canarias,
		Hamburger Sternwarte,
		Centro de Astrobiolog\'{\i}a and
		Centro Astron\'omico Hispano-Alem\'an), 
		with additional contributions by the MINECO, 
		the Deutsche Forschungsgemeinschaft through the Major Research Instrumentation Programme and Research Unit FOR2544 ``Blue Planets around Red Stars'', 
		the Klaus Tschira Stiftung, 
		the states of Baden-W\"urttemberg and Niedersachsen, 
		and by the Junta de Andaluc\'{\i}a.
		We acknowledge financial support from the Agencia Estatal de Investigaci\'on (AEI/10.13039/501100011033) of the Ministerio de Ciencia e Innovaci\'on and the ERDF ``A way of making Europe'' through projects 
		PID2022-137241NB-C4[1:4],	% CAB+IAA+IAC+UCM
		PID2021-125627OB-C31,		% ICE
		%				% Your active AYA/ESP project  
		PID2022-141216NB-I00, and the Centre of Excellence ``Severo Ochoa'' and ``Mar\'ia de Maeztu'' awards to the Instituto de Astrof\'isica de Canarias (CEX2019-000920-S), Instituto de Astrof\'isica de Andaluc\'ia (CEX2021-001131-S) and Institut de Ci\`encies de l'Espai (CEX2020-001058-M). F.L. and L.N. acknowledge the support by the Deutsche Forschungsgemeinschaft (DFG, German Research Foundation) – Project number 314665159. D.S. acknowledges financial support from the project PID2021-126365NB-C21(MCI/AEI/FEDER, UE)
		and from the Severo Ochoa grant CEX2021-001131-S funded by MCIN/AEI/10.13039/501100011033. D.C. is supported by the LMU-Munich Fraunhofer-Schwarzschild Fellowship and by the Deutsche Forschungsgemeinschaft (DFG, German Research Foundation) under Germany's Excellence Strategy – EXC 2094 – 390783311.
		
		%Based on data from the CARMENES data archive at CAB (CSIC-INTA).
	\end{acknowledgements}
	
	% Please note that we have included the references to the file aa. dem in
	\bibliographystyle{aa}
	\bibliography{ref}

	\begin{appendix}
		\section{Additional tables and figures}
		
		\begin{figure}
			\centering
			\includegraphics[width=9cm]{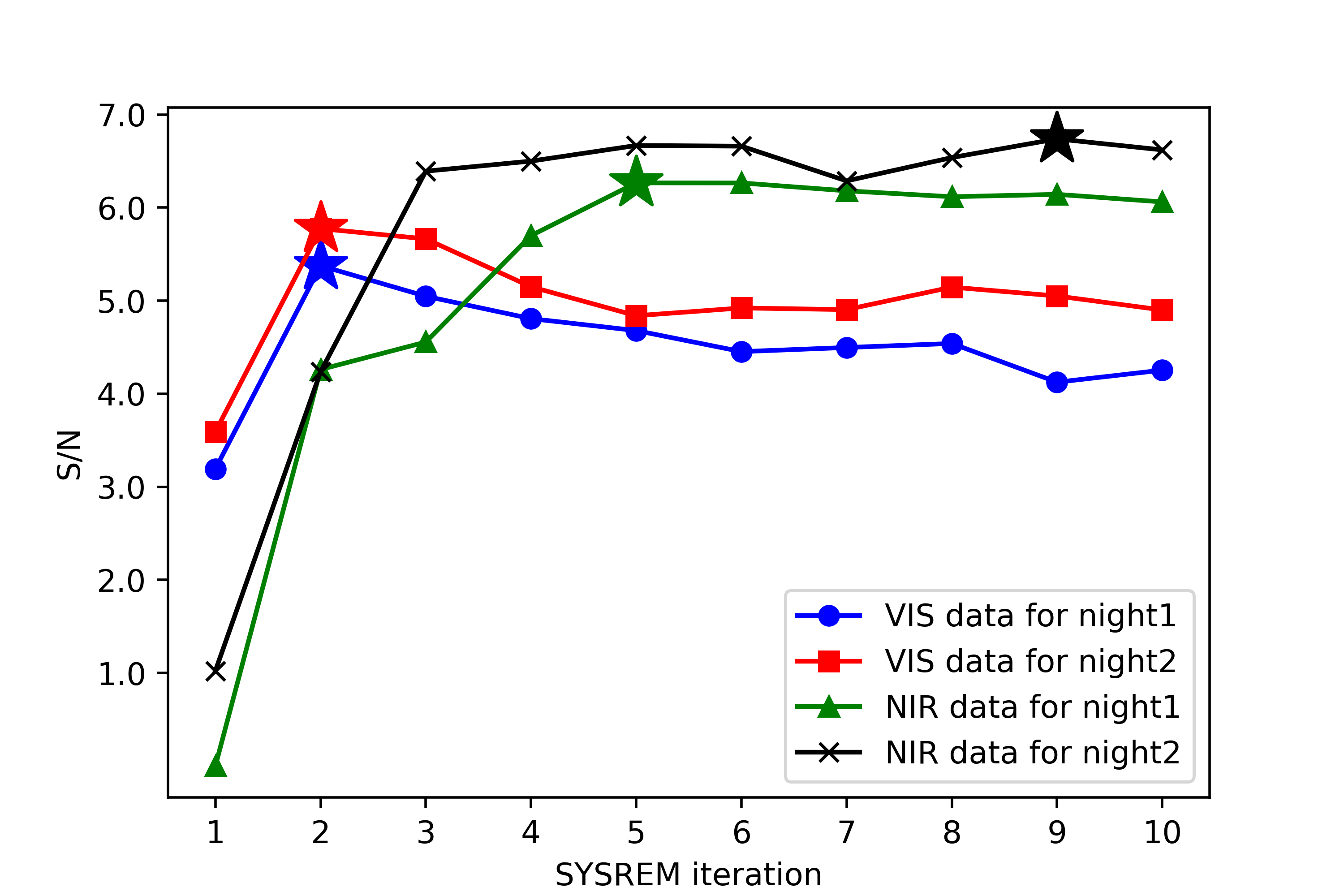}
			\caption{Corresponding S/N of \ion{Fe}{i} signal at different \texttt{SYSREM} iteration numbers. The maximum S/N values are marked with a star. These S/N values are measured with the peak location where the detection signal is the strongest.}
			\label{sysrem_iter_Fe}
		\end{figure}
		
		\begin{figure}
			\centering
			\includegraphics[width=9cm]{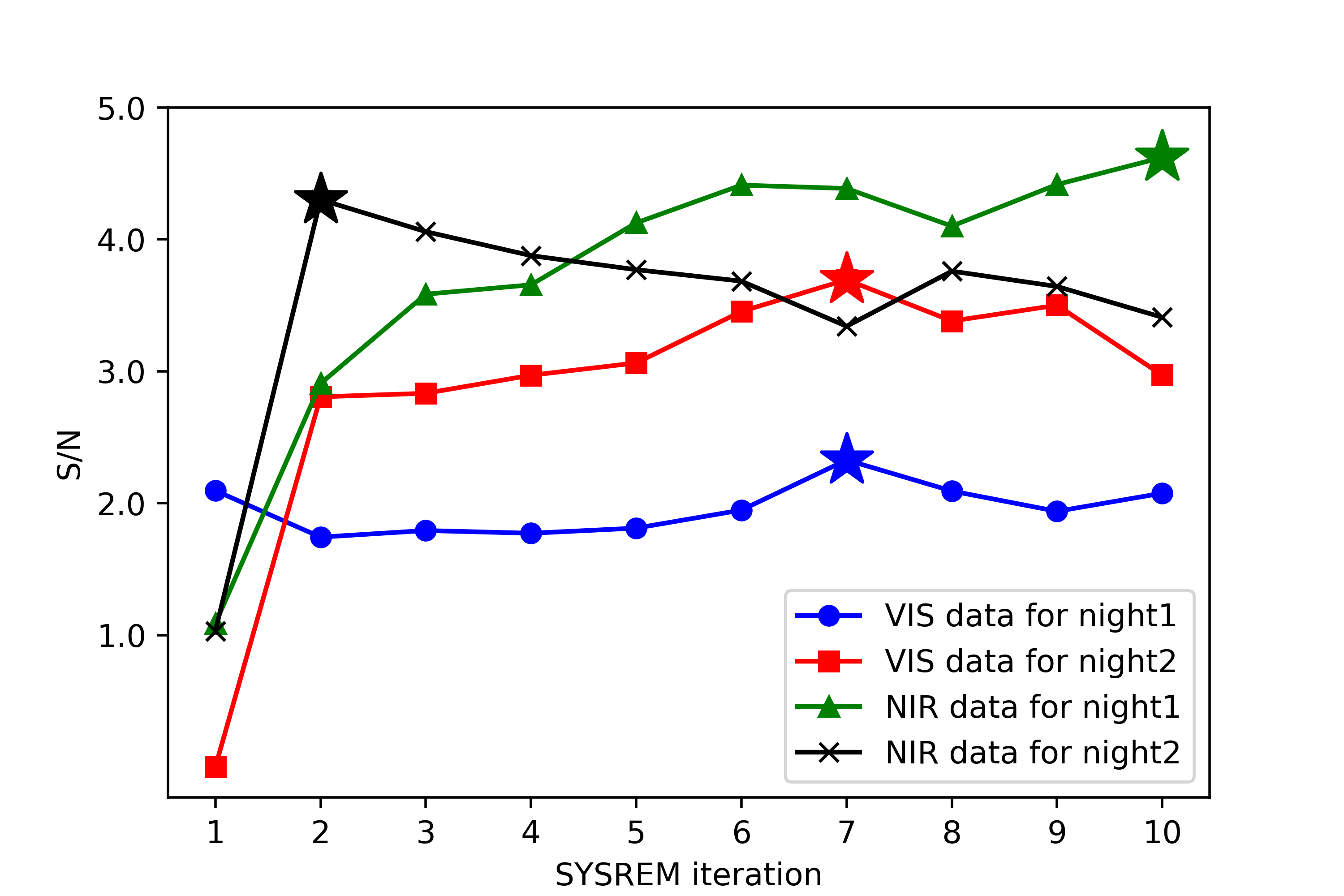}
			\caption{Same as Fig.\ref{sysrem_iter_Fe}, but for the \ion{Ti}{i} signal.}
			\label{sysrem_iter_Ti}
		\end{figure}
		
		\begin{figure*}
			\centering
			\includegraphics[width=17cm]{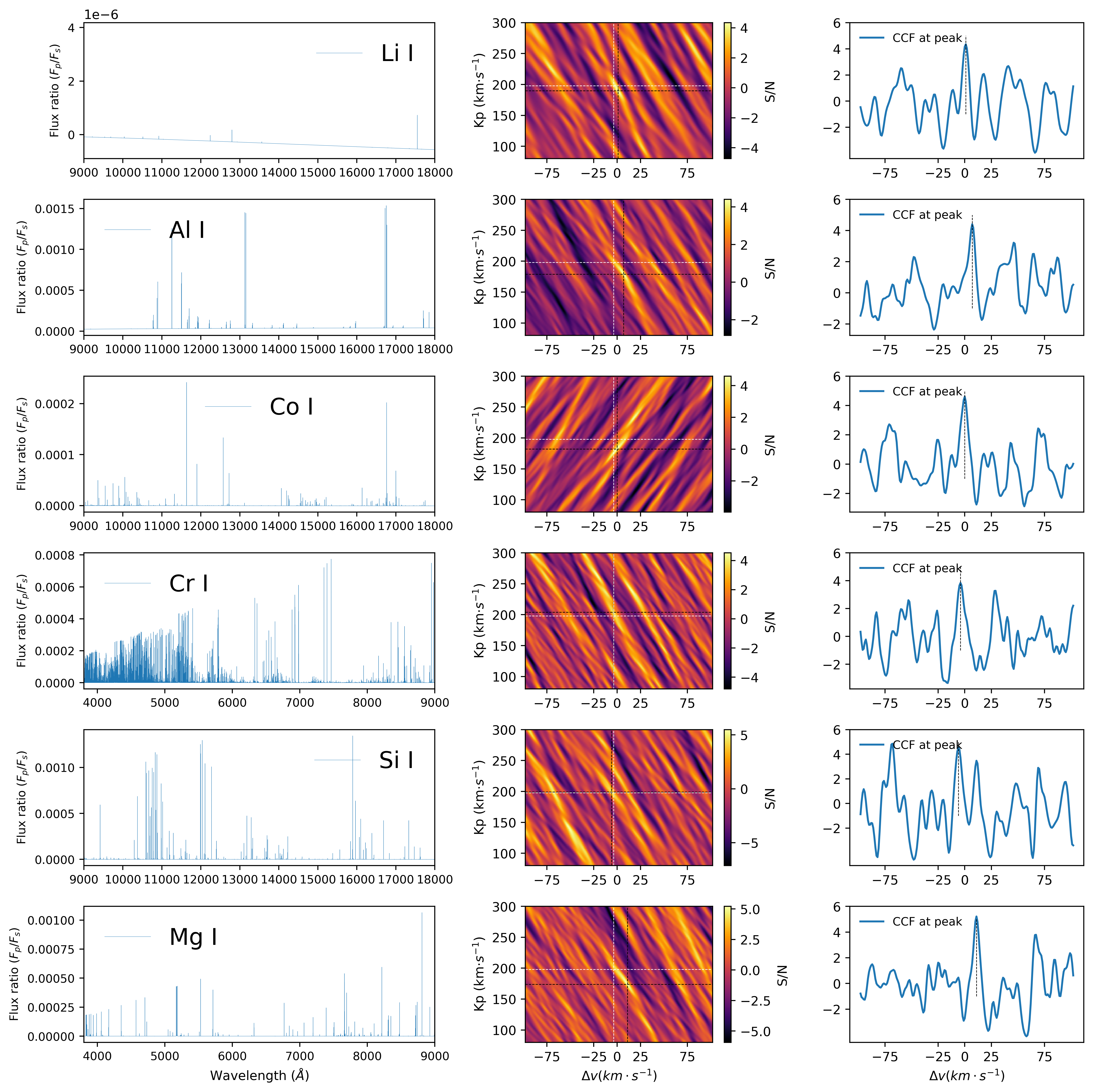}
			\caption{Some tentative signals for certain chemical species. \ion{Li}{i}, \ion{Al}{i}, \ion{Cr}{i}, and \ion{Si}{i} data were from the NIR spectra of night one, \ion{Co}{i} data was from the VIS spectra of night two, and \ion{Mg}{i} data was from the VIS spectra of night one. In the middle panel, the dashed white lines indicate the location of the maximum \ion{Fe}{i} signal, and the dashed black lines indicate the location where the corresponding chemical species detection significance is at its maximum.}
			\label{some_signal}
		\end{figure*}
		
		\begin{figure*}
			\centering
			\includegraphics[width=17cm]{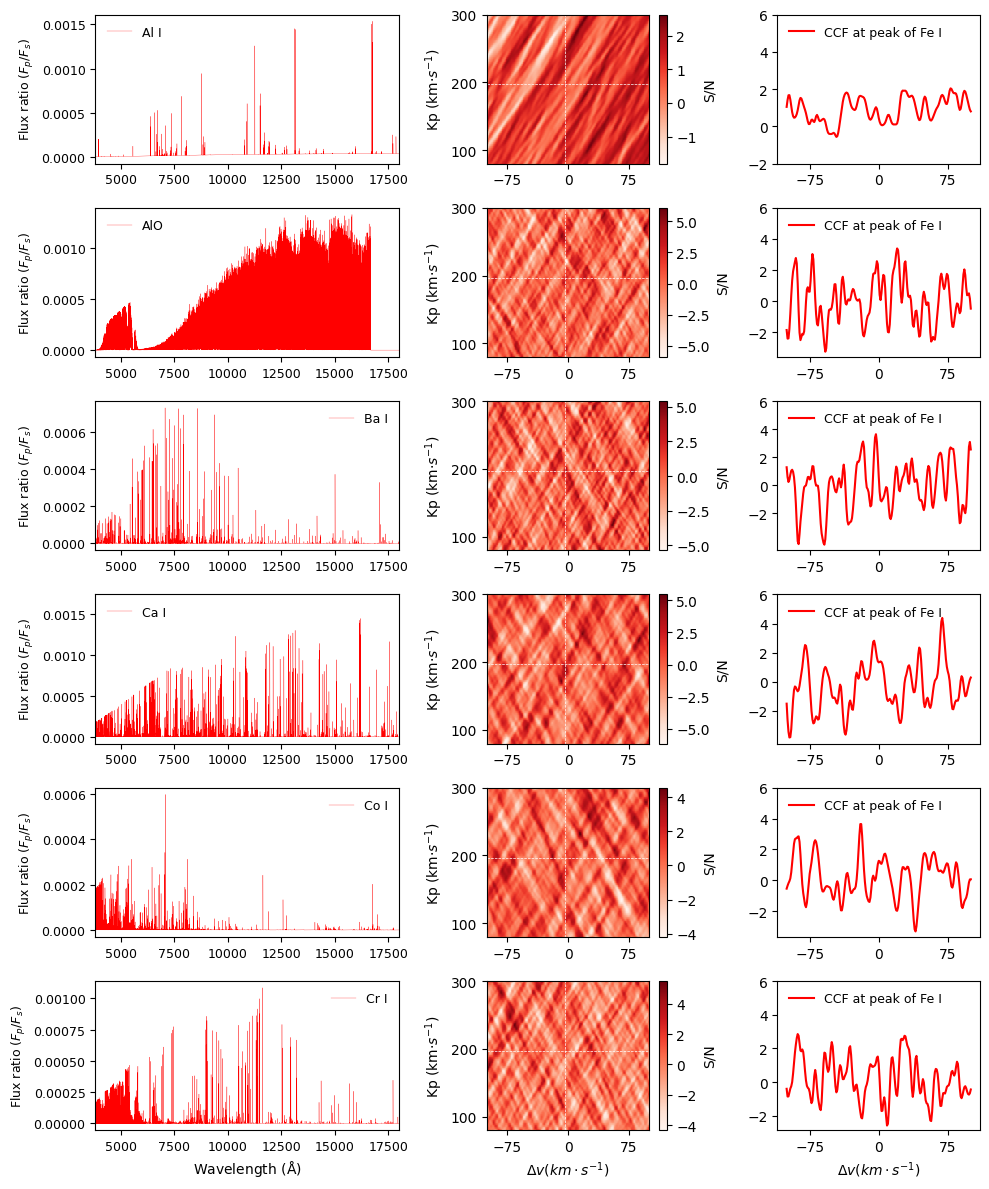}
			\caption{Non-detection of other chemical species. \textit{Left panels}: Spectral model of each species. These are normalized spectra that were calculated in a similar way to that described in Sect.\ref{model}. \textit{Middle panels}: Combined two-night S/N maps of each species. The dashed white lines indicate the peak location from the \ion{Fe}{i} signal. \textit{Right panels}: CCF at the peak of \ion{Fe}{i}.}
			\label{Kpmap_no_signal}
		\end{figure*}
		
		\begin{figure*}
			\centering
			\includegraphics[width=17cm]{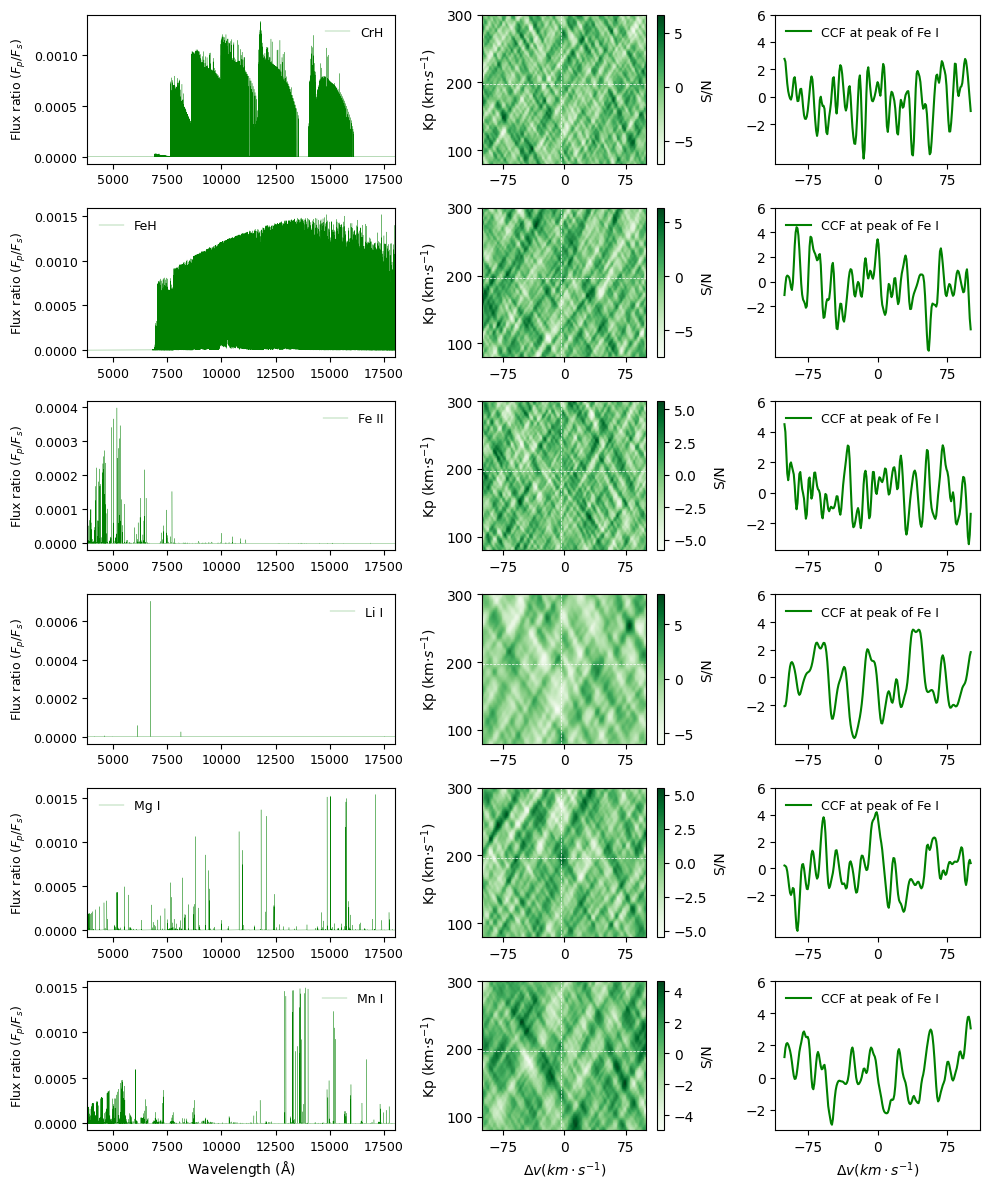}
			\caption{Same as Fig.\ref{Kpmap_no_signal}, but for different chemical species.}
			\label{Kpmap_no_signal2}
		\end{figure*}
		
		\begin{figure*}
			\centering
			\includegraphics[width=17cm]{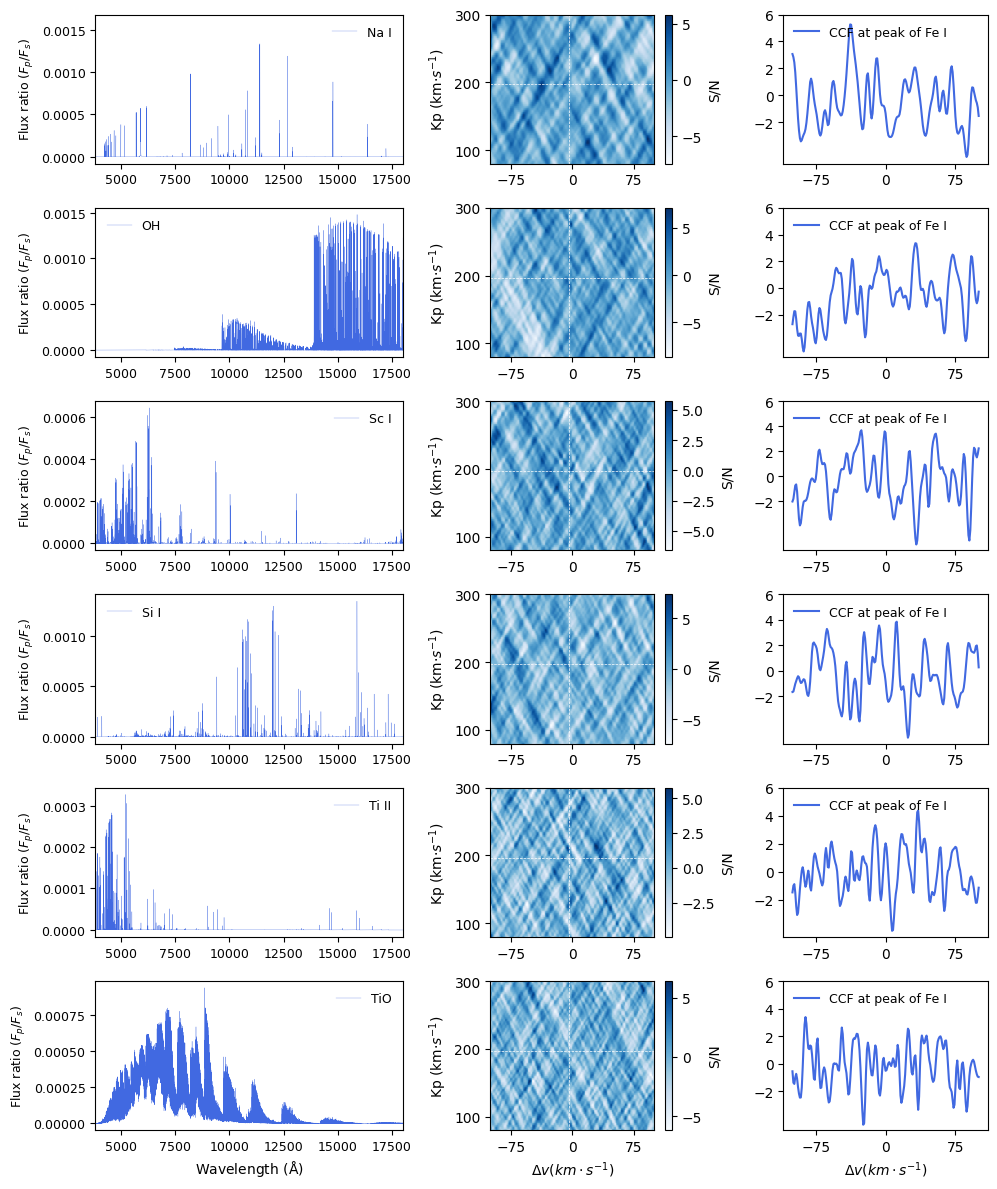}
			\caption{Same as Fig.\ref{Kpmap_no_signal}, but for different chemical species.}
			\label{Kpmap_no_signal3}
		\end{figure*}
		
		\begin{figure*}
			\centering
			\includegraphics[width=17cm]{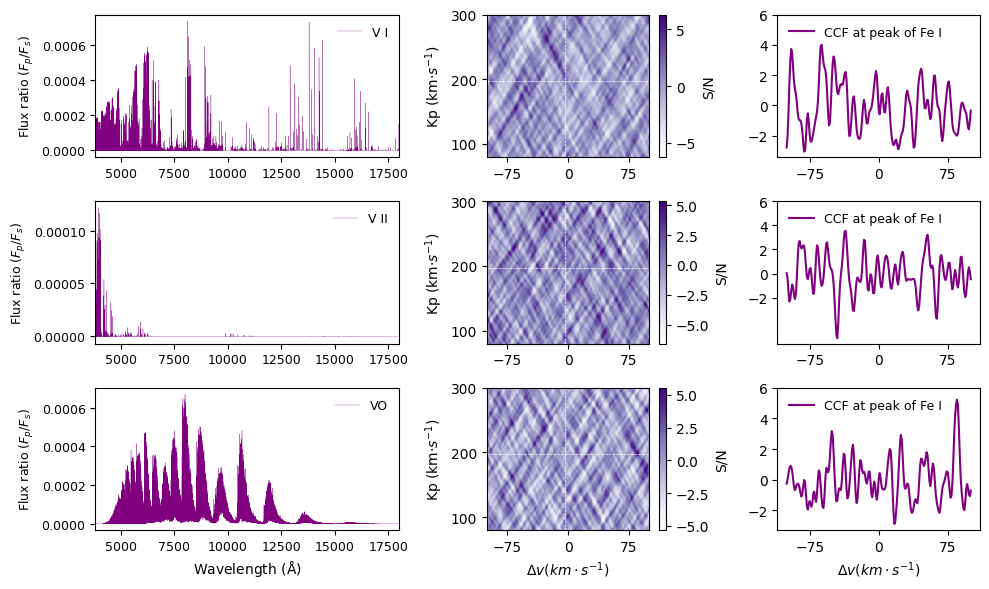}
			\caption{Same as Fig.\ref{Kpmap_no_signal}, but for different chemical species.}
			\label{Kpmap_no_signal4}
		\end{figure*}

		\begin{figure*}
			\centering
			\includegraphics[width=17cm]{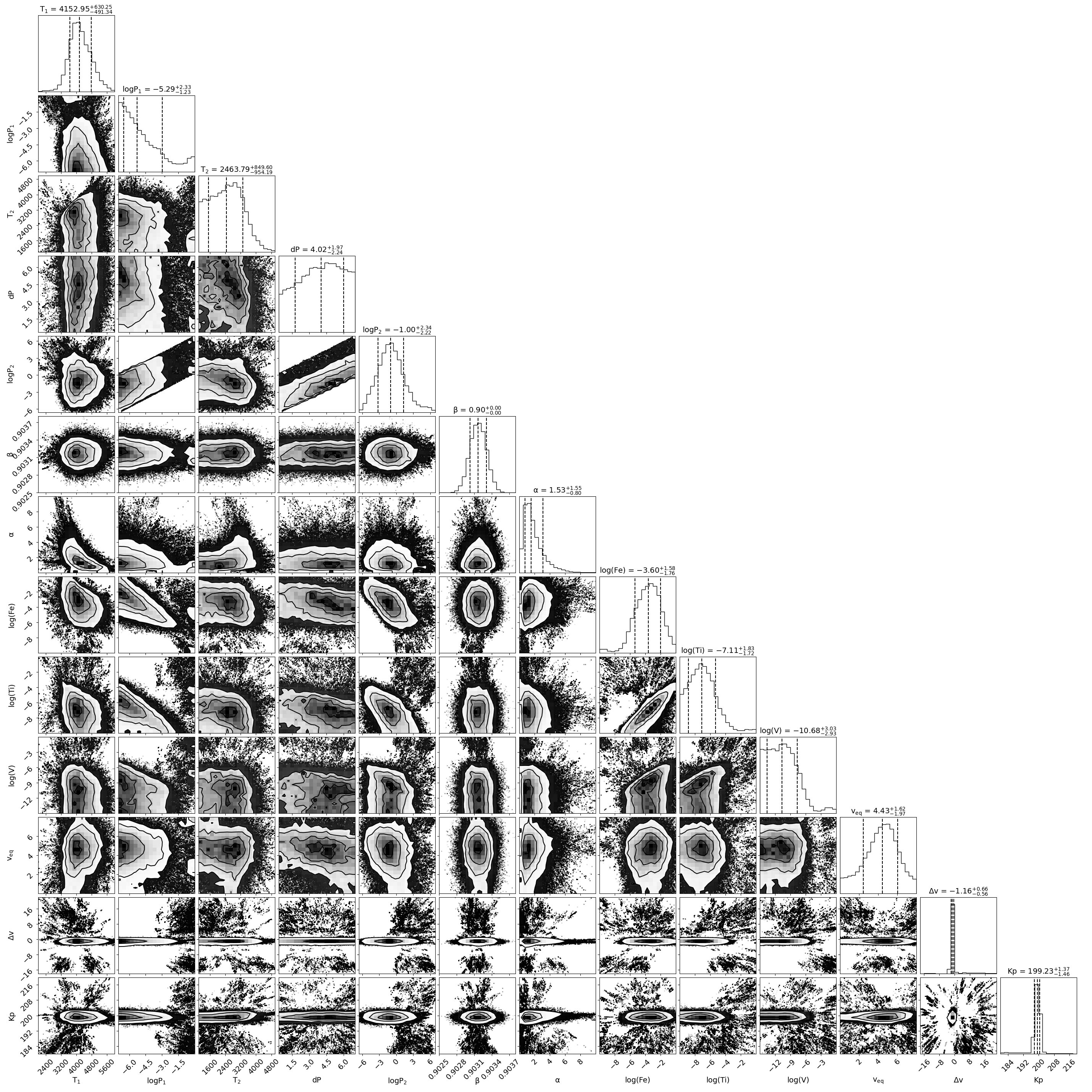}
			\caption{Posterior distribution of the parameters from the atmospheric retrieval. Here, the volume mixing ratios of \ion{Fe}{i}, \ion{Ti}{i}, and \ion{V}{i} are assumed to be constant throughout the atmospheric structure.}
			\label{corner_map}
		\end{figure*}

		\begin{figure*}
			\centering
			\includegraphics[width=17cm]{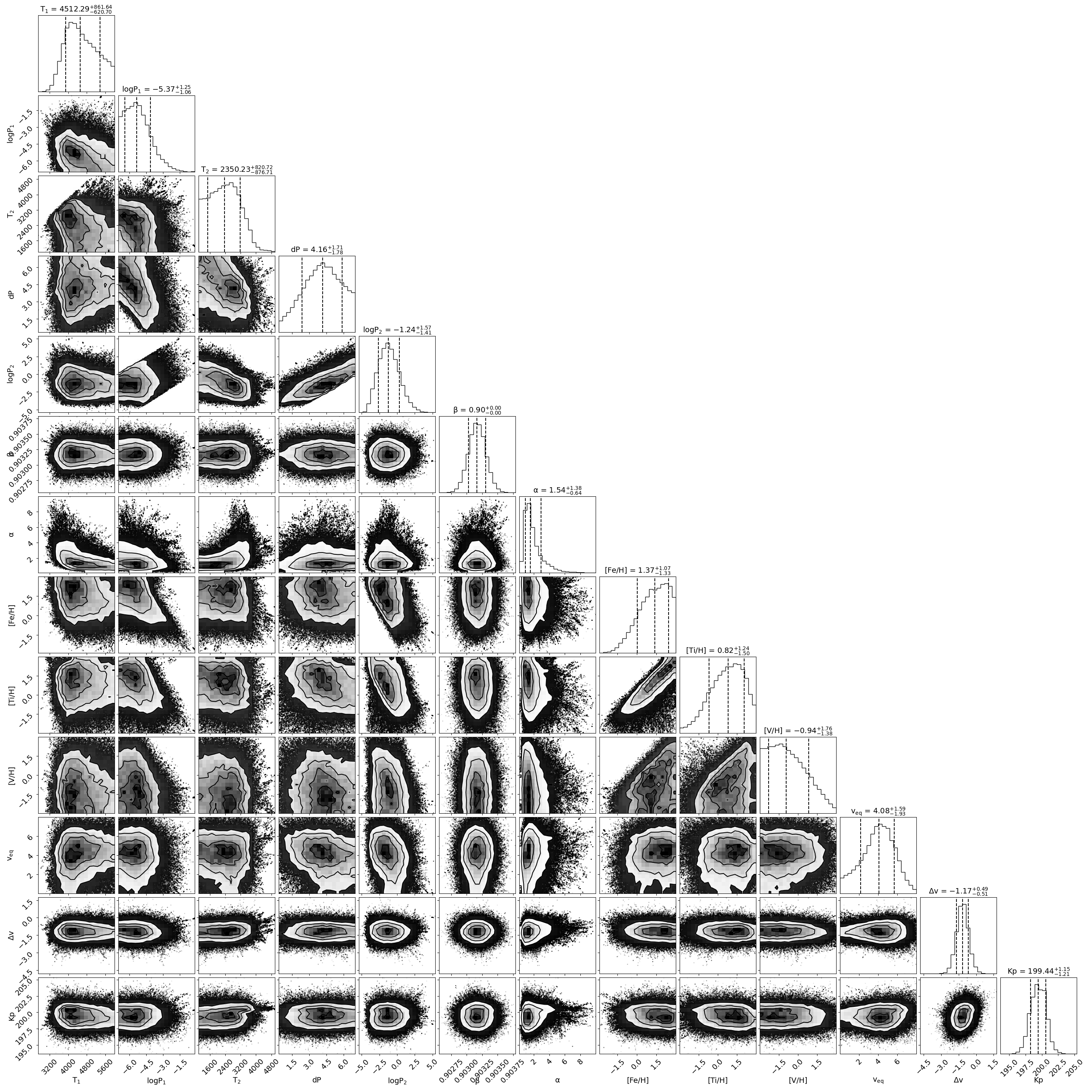}
			\caption{Same as Fig.\ref{corner_map}, but the volume mixing ratios of \ion{Fe}{i}, \ion{Ti}{i}, and \ion{V}{i} were calculated with the chemical equilibrium grid.}
			\label{corner_map2}
		\end{figure*}
		
	\end{appendix}

\end{document}